\documentclass[12pt,preprint]{aastex}
\usepackage{natbib}
\singlespace

\shorttitle{\WMAP\ 5-year Overview}
\shortauthors{Hinshaw et al.}

\newcommand{\map}    {{\sl WMAP}}

\newcommand{\dg}     {\mbox{$^{\circ}$}}

\newcommand{\lt}     {\mbox{$<$}}
\newcommand{\gt}     {\mbox{$>$}}

\newcommand{\beq}    {\begin{equation}}
\newcommand{\eeq}    {\end{equation}}
\newcommand{\beqa}   {\begin{eqnarray}}
\newcommand{\eeqa}   {\end{eqnarray}}

\newcommand{\fnlKS}  {f_{NL}^{\rm local}}
\newcommand{\fnleq}  {f_{NL}^{\rm equil}} 

\slugcomment{Astrophysical Journal Supplement Series, in press}

\begin{document}

\title{Five-Year Wilkinson Microwave Anisotropy Probe 
(WMAP\altaffilmark{1}) Observations:\\
Data Processing, Sky Maps, \& Basic Results}

\author{{G. Hinshaw} \altaffilmark{2},
{J. L. Weiland} \altaffilmark{3},
{R. S. Hill}    \altaffilmark{3},
{N. Odegard} \altaffilmark{3},
{D. Larson}  \altaffilmark{4},
{C. L. Bennett} \altaffilmark{4},
{J. Dunkley} \altaffilmark{5,6,7},
{B. Gold}    \altaffilmark{4},
{M. R. Greason} \altaffilmark{3},
{N. Jarosik} \altaffilmark{5},
{E. Komatsu} \altaffilmark{8},
{M. R. Nolta}   \altaffilmark{9},
{L. Page}    \altaffilmark{5},
{D. N. Spergel} \altaffilmark{6,10},
{E. Wollack} \altaffilmark{2},
{M. Halpern} \altaffilmark{11},
{A. Kogut}   \altaffilmark{2},
{M. Limon}   \altaffilmark{12}, 
{S. S. Meyer}   \altaffilmark{13},
{G. S. Tucker}  \altaffilmark{14},
{E. L. Wright}  \altaffilmark{15}}

\altaffiltext{1}{\map\ is the result of a partnership between Princeton 
                 University and NASA's Goddard Space Flight Center. Scientific 
		 guidance is provided by the \map\ Science Team.}
\altaffiltext{2}{{Code 665, NASA/Goddard Space Flight Center,  Greenbelt, MD 20771}}
\altaffiltext{3}{{Adnet Systems, Inc.,  7515 Mission Dr., Suite A100, Lanham, Maryland 20706}}
\altaffiltext{4}{{Dept. of Physics \& Astronomy,  The Johns Hopkins University, 3400 N. Charles St.,  Baltimore, MD  21218-2686}}
\altaffiltext{5}{{Dept. of Physics, Jadwin Hall,  Princeton University, Princeton, NJ 08544-0708}}
\altaffiltext{6}{{Dept. of Astrophysical Sciences,  Peyton Hall, Princeton University, Princeton, NJ 08544-1001}}
\altaffiltext{7}{{Astrophysics, University of Oxford,  Keble Road, Oxford, OX1 3RH, UK}}
\altaffiltext{8}{{Univ. of Texas, Austin, Dept. of Astronomy,  2511 Speedway, RLM 15.306, Austin, TX 78712}}
\altaffiltext{9}{{Canadian Institute for Theoretical Astrophysics,  60 St. George St, University of Toronto,  Toronto, ON  Canada M5S 3H8}}
\altaffiltext{10}{{Princeton Center for Theoretical Physics,  Princeton University, Princeton, NJ 08544}}
\altaffiltext{11}{{Dept. of Physics and Astronomy, University of  British Columbia, Vancouver, BC  Canada V6T 1Z1}}
\altaffiltext{12}{{Columbia Astrophysics Laboratory,  550 W. 120th St., Mail Code 5247, New York, NY  10027-6902}}
\altaffiltext{13}{{Depts. of Astrophysics and Physics, KICP and EFI,  University of Chicago, Chicago, IL 60637}}
\altaffiltext{14}{{Dept. of Physics, Brown University, 182 Hope St., Providence, RI 02912-1843}}
\altaffiltext{15}{{UCLA Physics \& Astronomy, PO Box 951547,  Los Angeles, CA 90095-1547}}

\email{Gary.F.Hinshaw@nasa.gov}

\begin{abstract}

We present new full-sky temperature and polarization maps in five frequency
bands from 23 to 94 GHz, based on data from the first five years of the \map\
sky survey.  The new maps are consistent with previous maps and are more
sensitive.  The five-year maps incorporate several improvements in data
processing made possible by the additional years of data and by a more complete
analysis of the instrument calibration and in-flight beam response.  We present
several new tests for systematic errors in the polarization data and conclude
that W band polarization data is not yet suitable for cosmological studies, but
we suggest directions for further study. We {\em do} find that Ka band data is
suitable for use; in conjunction with the additional years of data, the addition
of Ka band to the previously used Q and V band channels significantly reduces
the uncertainty in the optical depth parameter, $\tau$.  Further scientific
results from the five year data analysis are presented in six companion papers
and are summarized in \S\ref{sec:science} of this paper.  

With the 5 year \map\ data, we detect no convincing deviations from the minimal
6-parameter $\Lambda$CDM model: a flat universe dominated by a cosmological
constant, with adiabatic and nearly scale-invariant Gaussian fluctuations. 
Using \map\ data combined with measurements of Type Ia supernovae (SN) and
Baryon Acoustic Oscillations (BAO) in the galaxy distribution, we find (68\% CL
uncertainties):
\ensuremath{\Omega_bh^2 = 0.02267^{+ 0.00058}_{- 0.00059}},
\ensuremath{\Omega_ch^2 = 0.1131\pm 0.0034},
\ensuremath{\Omega_\Lambda = 0.726\pm 0.015},
\ensuremath{n_s = 0.960\pm 0.013},
\ensuremath{\tau = 0.084\pm 0.016}, and
\ensuremath{\Delta_{\cal R}^2 = (2.445\pm 0.096)\times 10^{-9}} at $k=0.002~{\rm Mpc^{-1}}$. 
From these we derive:
\ensuremath{\sigma_8 = 0.812\pm 0.026},
\ensuremath{H_0 = 70.5\pm 1.3}~${\rm
km~s^{-1}~Mpc^{-1}}$,
\ensuremath{\Omega_b = 0.0456\pm 0.0015},
\ensuremath{\Omega_c = 0.228\pm 0.013},
\ensuremath{\Omega_mh^2 = 0.1358^{+ 0.0037}_{- 0.0036}},
\ensuremath{z_{\rm reion} = 10.9\pm 1.4},
and 
\ensuremath{t_0 = 13.72\pm 0.12\ \mbox{Gyr}}.
The new limit on the tensor-to-scalar ratio is
\ensuremath{r < 0.22\ \mbox{(95\% CL)}}, 
while the evidence for a running spectral index is insignificant,
\ensuremath{dn_s/d\ln{k} = -0.028\pm 0.020} (68\% CL).
We obtain tight, simultaneous limits on the (constant)
dark energy equation of state and the spatial curvature of the
universe: 
\ensuremath{-0.14<1+w<0.12\ \mbox{(95\% CL)}}
and
\ensuremath{-0.0179<\Omega_k<0.0081\ \mbox{(95\% CL)}}. 
The number of relativistic degrees of freedom, expressed in units of the
effective number of neutrino species, is found to be
\ensuremath{N_{\rm eff} = 4.4\pm 1.5} (68\% CL), 
consistent with the standard value of 3.04. Models with $N_{\rm eff} = 0$ are
disfavored at $\gt$99.5\% confidence.  Finally, new limits on physically
motivated primordial non-Gaussianity parameters are $-9 < \fnlKS <111$ (95\% CL)
and $-151 < \fnleq < 253$ (95\% CL) for the local and equilateral models,
respectively.

\end{abstract}


\section{INTRODUCTION}\label{sec:intro}

The Wilkinson Microwave Anisotropy Probe (\map) is a Medium-Class Explorer
(MIDEX) satellite aimed at elucidating cosmology through full-sky observations
of the cosmic microwave background (CMB).  The \map\ full-sky maps of the
temperature and polarization anisotropy in five frequency bands provide our most
accurate view to date of conditions in the early universe.  The multi-frequency
data facilitate the separation of the CMB signal from foreground emission
arising both from our Galaxy and from extragalactic sources.  The CMB angular
power spectrum derived from these maps exhibits a highly coherent acoustic peak
structure which makes it possible to extract a wealth of information about the
composition and history of the universe, as well as the processes that seeded
the fluctuations.

\map\ data \citep{bennett/etal:2003, spergel/etal:2003, hinshaw/etal:2007,
spergel/etal:2007}, along with a host of pioneering CMB experiments
\citep{miller/etal:1999, lee/etal:2001, netterfield/etal:2002,
halverson/etal:2002, pearson/etal:2003, scott/etal:2003, benoit/etal:2003b}, and
other cosmological measurements \citep{percival/etal:2001, tegmark/etal:2004a,
cole/etal:2005, tegmark/etal:2006, eisenstein/etal:2005, percival/etal:2007c,
astier/etal:2006, riess/etal:2007, wood-vasey/etal:2007} have established
$\Lambda$CDM as the standard model of cosmology: a flat universe dominated by
dark energy, supplemented by dark matter and atoms with density fluctuations
seeded by a Gaussian, adiabatic, nearly scale invariant process.  The basic
properties of this universe are determined by five numbers: the density of
matter, the density of atoms, the age of the universe (or equivalently, the
Hubble constant today), the amplitude of the initial fluctuations, and their
scale dependence.

By accurately measuring the first few peaks in the angular power spectrum and
the large-scale polarization anisotropy, \map\ data have enabled the following
inferences: 
\begin{itemize}

\item A precise (3\%) determination of the density of atoms in the universe. The
agreement between the atomic density derived from \map\ and the density inferred
from the deuterium abundance is an important test of the standard big bang
model.

\item A precise (3\%) determination of the dark matter density.  (With five
years of data and a better determination of our beam response, this measurement
has improved significantly.)  Previous CMB measurements have shown that the dark
matter must be non-baryonic and interact only weakly with atoms and radiation. 
The \map\ measurement of the density puts important constraints on
supersymmetric dark matter models and on the properties of other dark matter
candidates. 

\item A definitive determination of the acoustic scale at redshift $z=1090$. 
Similarly, the recent measurement of baryon acoustic oscillations (BAO) in the
galaxy power spectrum \citep{eisenstein/etal:2005} has determined the acoustic
scale at redshift $z \sim 0.35$.  When combined, these standard rulers
accurately measure the geometry of the universe and the properties of the dark
energy.  These data  require a nearly flat universe dominated by dark energy
consistent with a cosmological constant.

\item A precise determination of the Hubble Constant, in conjunction with BAO
observations.  Even when allowing curvature ($\Omega_0 \ne 1$) and a free dark
energy equation of state ($w \ne -1$), the acoustic data determine the Hubble
constant to within 3\%.  The measured value is in excellent agreement with
independent results from the Hubble Key Project \citep{freedman/etal:2001},
providing yet another important consistency test for the standard model.

\item Significant constraint of the basic properties of the primordial
fluctuations.  The anti-correlation seen in the temperature/polarization (TE)
correlation spectrum on 4$^{\circ}$ scales implies that the fluctuations are
primarily adiabatic and rule out defect models and isocurvature models as the
primary source of fluctuations \citep{peiris/etal:2003}.
\end{itemize}

Further, the \map\ measurement of the primordial power spectrum of matter
fluctuations constrains the physics of inflation, our best model for the origin
of these fluctuations.  Specifically, the 5 year data provide the best
measurement to date of the scalar spectrum's amplitude and slope, and place the
most stringent limits to date on the amplitude of tensor fluctuations.  However,
it should be noted that these constraints assume a smooth function of scale,
$k$.  Certain models with localized structure in $P(k)$, and hence additional
parameters, are not ruled out, but neither are they required by the data; see
e.g. \citet{shafieloo/souradeep:2007, hunt/sarkar:2007}.

The statistical properties of the CMB fluctuations measured by \map\ are close
to Gaussian; however, there are several hints of possible deviations from
Gaussianity, e.g. \citet{eriksen/etal:2007c, copi/etal:2007,
land/magueijo:2007, yadav/wandelt:2008}.  Significant deviations would be a
very important signature of new physics in the early universe.  

Large-angular-scale polarization measurements currently provide our best window
into the universe at $z \sim 10$.  The \map\ data imply that the universe was
reionized long before the epoch of the oldest known quasars.  By accurately
constraining the optical depth of the universe, \map\ not only constrains the
age of the first stars but also determines the amplitude of primordial
fluctuations to better than 3\%.  This result is important for
constraining the growth rate of structure.

This paper summarizes results compiled from 5 years of \map\ data that are fully
presented in a suite of 7 papers (including this one).  The new results improve
upon previous results in many ways: additional data reduces the random noise,
which is especially important for studying the temperature signal on small
angular scales and the polarization signal on large angular scales; five
independent years of data enable comparisons and null tests that were not
previously possible; the instrument calibration and beam response have been much
better characterized, due in part to improved analyses and to additional years
of data; and, other cosmological data have become available.

In addition to summarizing the other papers, this paper reports on changes in
the \map\ data processing pipeline, presents the 5 year temperature and
polarization maps, and gives new results on instrument calibration and on
potential systematic errors in the polarization data. \citet{hill/etal:prep}
discuss the program to derive an improved physical optics model of the \map\
telescope, and use the results to better determine the \map\ beam response. 
\citet{gold/etal:prep} present a new analysis of diffuse foreground emission in
the \map\ data and update previous analyses using 5 year data. 
\citet{wright/etal:prep} analyze extragalactic point sources and provide an
updated source catalog, with new results on source variability. 
\citet{nolta/etal:prep} derive the angular power spectra from the maps,
including the TT, TE, TB, EE, EB, and BB spectra.  \citet{dunkley/etal:prep}
produce an updated likelihood function and present cosmological parameter
results based on 5 year \map\ data.  They also develop an independent analysis
of polarized foregrounds and use those results to test the reliability of the
optical depth inference to foreground removal errors.  \citet{komatsu/etal:prep}
infer cosmological parameters by combining 5 year \map\ data with a host of
other cosmological data and discuss the implications of the results.  Concurrent
with the submission of these papers, all 5 year \map\ data are made available to
the research community via NASA's Legacy Archive for Microwave Background Data
Analysis (LAMBDA).  The data products are described in detail in the \map\
Explanatory Supplement \citep{limon/etal:prep}, also available on LAMBDA.

The \map\ instrument is composed of 10 differencing assemblies (DAs) spanning 5
frequencies from 23 to 94 GHz \citep{bennett/etal:2003}: 1 DA each at 23 GHz
(K1) and 33 GHz (Ka1), 2 each at 41 GHz (Q1,Q2) and 61 GHz (V1,V2), and 4 at 94
GHz (W1-W4).  Each DA is formed from two differential radiometers which are
sensitive to  orthogonal linear polarization modes; the radiometers are
designated  1 or 2  (e.g., V11 or W12) depending on polarization mode.

In this paper we follow the notation convention that flux density is  $S\sim
\nu^\alpha$ and antenna temperature is $T\sim \nu^\beta$, where the spectral 
indices are related by $\beta=\alpha-2$.   In general, the CMB is expressed in
terms of thermodynamic temperature, while Galactic and extragalactic foregrounds
are expressed in antenna temperature.  Thermodynamic temperature  differences
are given by $\Delta T = \Delta T_A [(e^x-1)^2/x^2e^x]$, where $x=h\nu/kT_0$, 
$h$ is the Planck constant, $\nu$ is the frequency, $k$ is the Boltzmann
constant, and  $T_0=2.725$ K is the CMB temperature \citep{mather/etal:1999}.  
A \map\ band-by-band tabulation of the conversion factors between thermodynamic
and  antenna temperature is given in Table~\ref{tab:radiometers}.

\section{CHANGES IN THE 5 YEAR DATA ANALYSIS}
\label{sec:change}

The 1 year and 3 year data analyses were described in detail in previous
papers.  In large part, the 5 year analysis employs the same methods, so we do
not repeat a detailed processing description here.  However, we have made
several improvements that are summarized here and described in more detail later
in this paper and in a series of companion papers, as noted.  We list the
changes in the order they appear in the processing pipeline:

\begin{itemize}

\item There is a $\sim 1'$ temperature-dependent pointing offset between the
star tracker coordinate system (which defines spacecraft coordinates) and the
instrument boresights.  In the 3 year analysis we introduced a correction to
account for the elevation change of the instrument boresights in spacecraft
coordinates.  With additional years of data, we have been able to refine our
thermal model of the pointing offset, so we now include a small ($\lt 1'$)
correction to account for the azimuth change of the instrument boresights. 
Details of the new correction are given in the 5 year Explanatory Supplement
\citep{limon/etal:prep}.

\item We have critically re-examined the relative and absolute intensity
calibration procedures, paying special attention to the absolute gain recovery 
obtainable from the modulation of the CMB dipole due to \map's motion.  We
describe the revised procedure in \S\ref{sec:cal_improve} and note that the sky
map calibration uncertainty has decreased from 0.5\% to 0.2\%.

\item The \map\ beam response has now been measured in 10 independent
``seasons'' of Jupiter observations.  In the highest resolution W band channels,
these measurements now probe the beam response $\sim$44 dB down from the beam
peak.  However, there is still non-negligible beam solid angle below this level
($\sim$0.5\%) that needs to be measured to enable accurate cosmological
inference. In the 3 year analysis we produced a physical optics model of the
A-side beam response starting with a pre-flight model and fitting in-flight
mirror distortions to the flight Jupiter data.  In the 5 year analysis we have
extended the model to the B-side optics and, for both sides, we have extended
the fit to include distortion modes a factor of 2 smaller in linear scale (4
times as many modes).  The model is used to augment the flight beam maps below a
given threshold.  The details of this work are given in \citet{hill/etal:prep}.

\item The far sidelobe response of the beam was determined from a combination of
ground measurements and in-flight lunar data taken early in the mission
\citep{barnes/etal:2003}.  For the current analysis, we have replaced a small
fraction of the far sidelobe data with the physical optics model described
above.  We have also made the following changes in our handling of the far
sidelobe pickup \citep{hill/etal:prep}: 1) We have enlarged the ``transition
radius'' that defines the boundary between the main beam and the far sidelobe
response.  This places a larger fraction of the total beam solid angle in the
main beam where uncertainties are easier to quantify and propagate into the
angular power spectra.  2) We have moved the far sidelobe deconvolution into the
combined calibration and sky map solver (\S\ref{sec:cal_improve}).  This
produces a self-consistent estimate of the intensity calibration and the
deconvolved sky map.  The calibrated time-ordered data archive has had an
estimate of the far sidelobe response subtracted from each datum (as it had in
the 3 year processing).

\item We have updated the optimal filters used in the final step of map-making. 
The functional form of the filter is unchanged \citep{jarosik/etal:2007}, but
the fits have been updated to cover years 4 and 5 of the flight data.

\item Each \map\ differencing assembly consists of two radiometers that are
sensitive to orthogonal linear polarization states.  The sum and difference of
the two radiometer channels split the signal into intensity and polarization
components, respectively.  However, the noise levels in the two radiometers are
not equal, in general, so more optimal sky map estimation is possible in theory,
at the cost of mixing intensity and polarization components in the process.  For
the current analysis, we investigated one such weighted algorithm and found that
the polarization maps were subject to unacceptable contamination by the
intensity signal in cases where the beam response was non-circular and the
gradient of the intensity signal was large, e.g., in K band.  As a result, we
reverted to the unweighted (and unbiased) estimator used in previous work.

\item We have improved the sky masks used to reject foreground contamination. 
In previous work, we defined masks based on contours of the K band data. In the
5 year analysis we produce masks based jointly on K band and Q band contours. 
For a given sky cut fraction, the new masks exclude flat spectrum (e.g.
free-free) emission more effectively.  The new masks are described in detail in
\citet{gold/etal:prep} and are provided with the 5 year data release.  In
addition, we have modified the ``processing'' mask used to exclude very bright
sources during sky map estimation.  The new mask is defined in terms of
low-resolution (r4) HEALPix sky pixels \citep{gorski/etal:2005} to facilitate a
cleaner definition of the pixel-pixel inverse covariance matrices, $N^{-1}$. 
One side effect of this change is to introduce a few r4-sized holes around the
brightest radio sources in the analysis mask, which incorporates the processing
mask as a subset.

\item We have amended our foreground analysis in the following ways: 1)
\citet{gold/etal:prep} perform a pixel-by-pixel analysis of the joint
temperature and polarization data to study the breakdown of the Galactic
emission into physical components.  2) We have updated some aspects of the
Maximum Entropy (MEM) based analysis, as described in \citet{gold/etal:prep}. 
3) \citet{dunkley/etal:prep} develop a new analysis of polarized foreground
emission using a Gibbs sampling approach that yields a cleaned CMB polarization
map and an associated covariance matrix.  4) \citet{wright/etal:prep} update the
\map\ point source catalog and present some results on variable sources in the 5
year data.  However, the basic cosmological results are still based on maps that
were cleaned with the same template-based procedure that was used in the 3 year
analysis.

\item We have improved the final temperature power spectrum, $C_l^{TT}$, by
using a Gibbs-based maximum likelihood estimate for $l \le 32$
\citep{dunkley/etal:prep} and a pseudo-$C_l$ estimate for higher $l$
\citep{nolta/etal:prep}.  As with the 3 year analysis, the pseudo-$C_l$ estimate
uses only V- and W-band data.  With 5 individual years of data and six V- and
W-band differencing assemblies, we can now form individual cross-power spectra
from 15 DA pairs within each of 5 years and from 36 DA pairs across 10 year
pairs, for a total of 435 independent cross-power spectra.  

\item In the 3 year analysis we developed a pseudo-$C_l$ method for evaluating
polarization power spectra in the presence of correlated noise.  In the present
analysis we additionally estimate the TE, TB, EE, EB, \& BB spectra and their
errors using an extension of the maximum likelihood method in
\citet{page/etal:2007}.  However, as in the 3 year analysis, the likelihood of a
given model is still evaluated directly from the polarization maps using a
pixel-based likelihood.

\item We have improved the form of the likelihood function used to infer
cosmological parameters from the Monte Carlo Markov Chains
\citep{dunkley/etal:prep}.  We use an exact maximum likelihood form for the $l
\le 32$ TT data \citep{eriksen/etal:2007}.  We have investigated theoretically
optimal methods for incorporating window function uncertainties into the
likelihood, but in tests with simulated data we have found them to be biased. 
In the end, we adopt the form used in the 3 year analysis
\citep{hinshaw/etal:2007}, but we incorporate the smaller 5 year window function
uncertainties \citep{hill/etal:prep} as inputs.  We now routinely account for
gravitational lensing when assessing parameters, and we have added an option to
use low-$l$ TB and EB data for testing non-standard cosmological models.

\item For testing nongaussianity, we employ an improved estimator for $f_{NL}$
\citep{creminelli/etal:2006, yadav/etal:prep}.  The results of this analysis are
described in \citet{komatsu/etal:prep}.

\end{itemize}

\section{OBSERVATIONS AND MAPS}
\label{sec:maps}

The 5 year \map\ data encompass the period from 00:00:00 UT, 10 August 2001 (day
number 222) to 00:00:00 UT, 9 August 2006 (day number 222).  The observing
efficiency during this time is roughly 99\%; Table~\ref{tab:baddata} lists the
fraction of data that was lost or rejected as unusable.  The Table also gives
the fraction of data that is flagged due to potential contamination by thermal
emission from Mars, Jupiter, Saturn, Uranus, and Neptune.  These data are not
used in map-making, but are useful for in-flight beam mapping
\citep{hill/etal:prep, limon/etal:prep}.

\begin{figure}
\epsscale{1.0}
\plotone{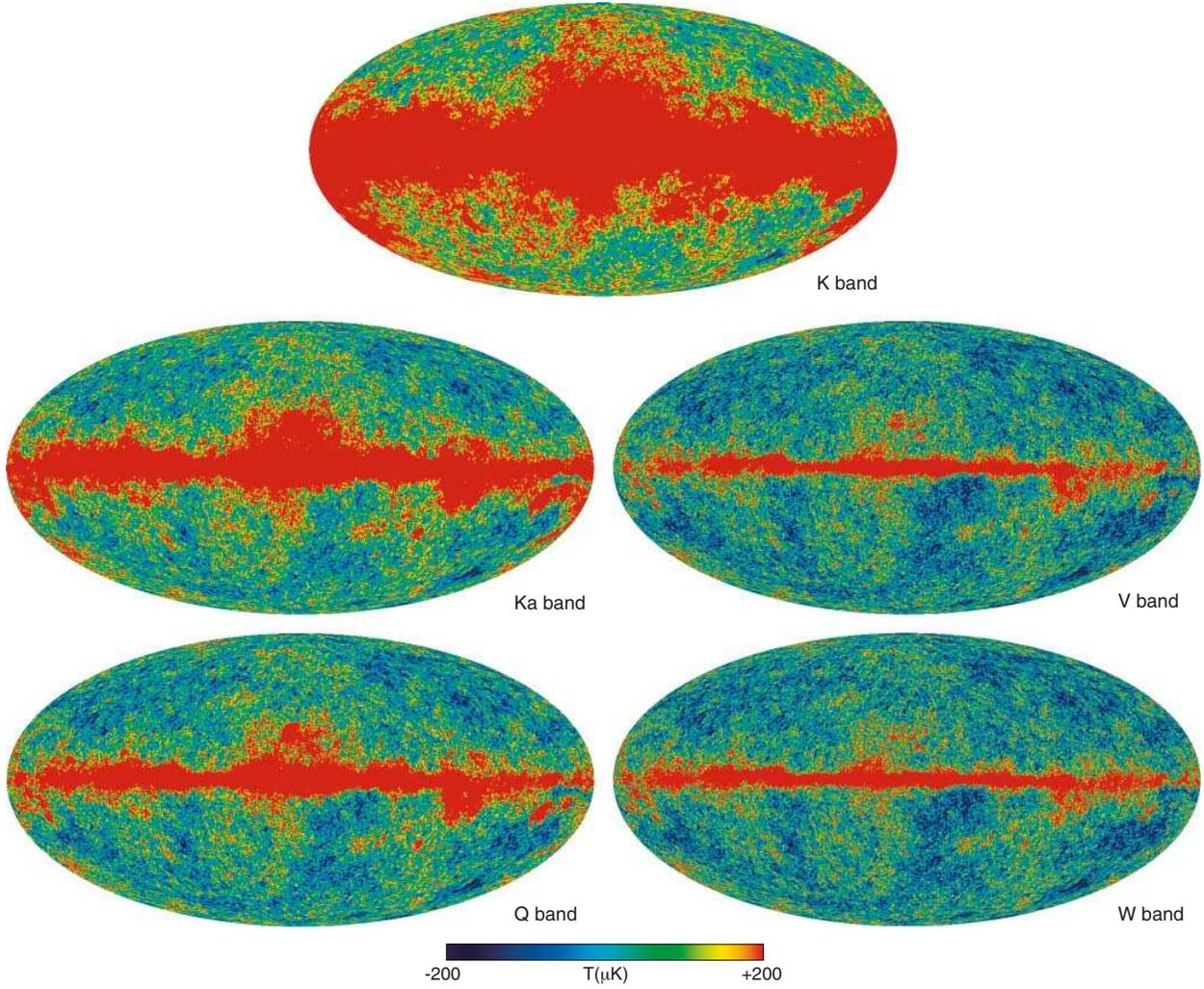}
\caption{Five-year temperature sky maps in Galactic coordinates smoothed with a
$0.2\dg$ Gaussian beam, shown in Mollweide projection. {\it top}: K band (23
GHz), {\it middle-left}: Ka band (33 GHz), {\it bottom-left}: Q band (41 GHz),
{\it middle-right}: V band (61 GHz), {\it bottom-right}: W band (94 GHz).}
\label{fig:i_maps}
\end{figure}

\begin{figure}
\epsscale{1.0}
\plotone{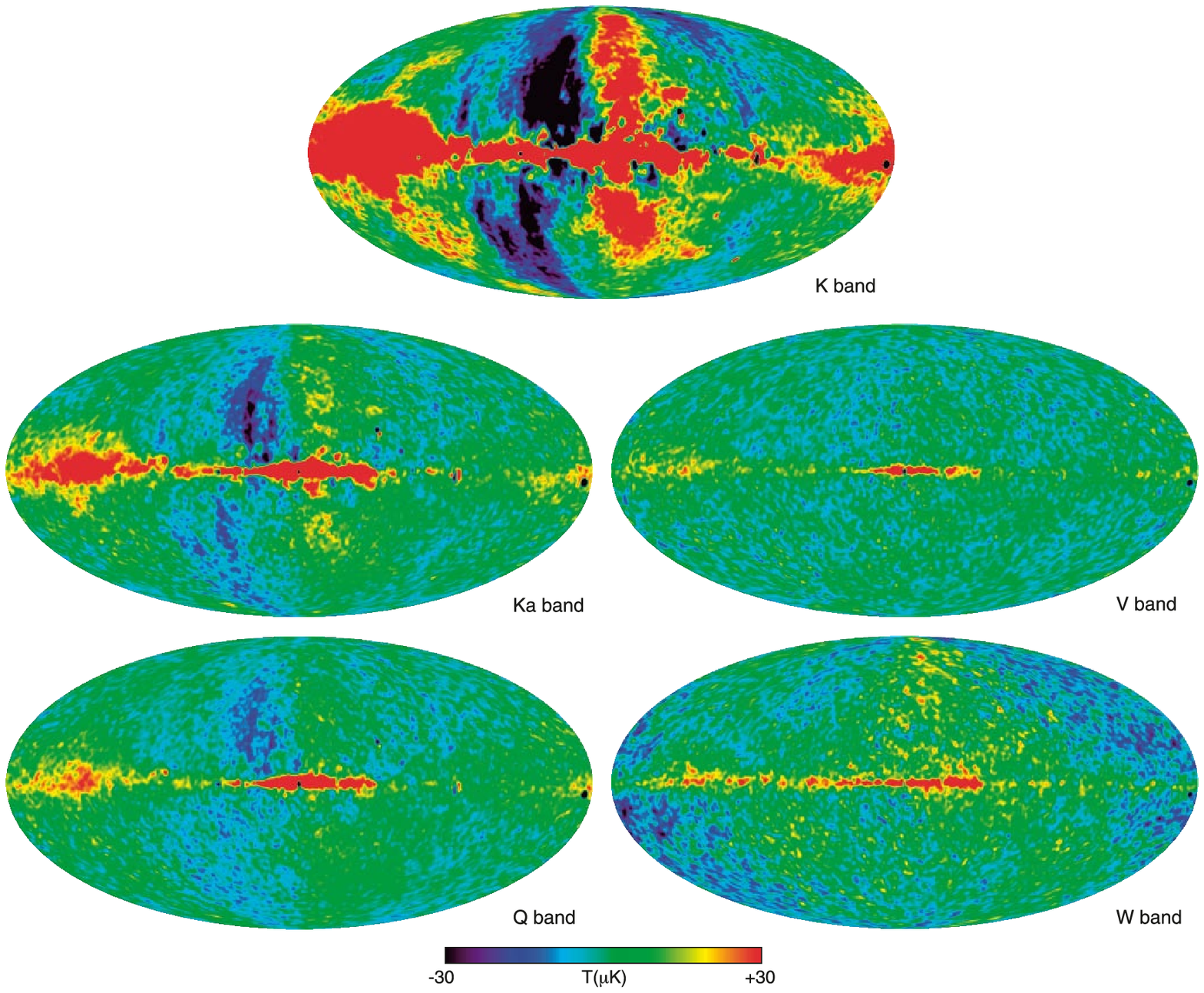}
\caption{Five-year Stokes Q polarization sky maps in Galactic coordinates
smoothed to an effective Gaussian beam of $2.0\dg$, shown in Mollweide
projection. {\it top}: K band (23 GHz), {\it middle-left}: Ka band (33 GHz),
{\it bottom-left}: Q band (41 GHz), {\it middle-right}: V band (61 GHz), {\it
bottom-right}: W band (94 GHz).}
\label{fig:q_maps}
\end{figure}

\begin{figure}
\epsscale{1.0}
\plotone{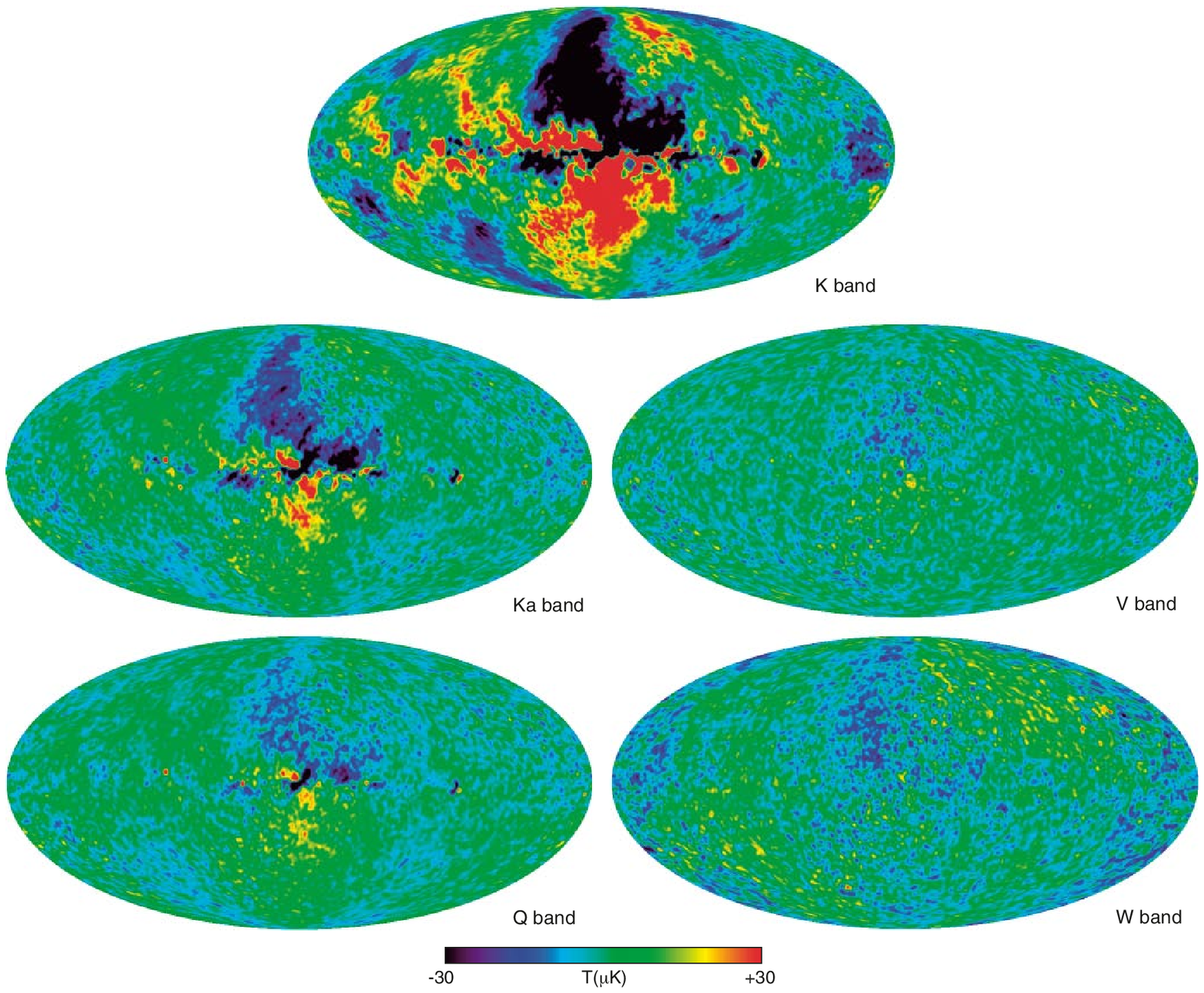}
\caption{Five-year Stokes U polarization sky maps in Galactic coordinates
smoothed to an effective Gaussian beam of $2.0\dg$, shown in Mollweide
projection. {\it top}: K band (23 GHz), {\it middle-left}: Ka band (33 GHz),
{\it bottom-left}: Q band (41 GHz), {\it middle-right}: V band (61 GHz), {\it
bottom-right}: W band (94 GHz).}
\label{fig:u_maps}
\end{figure}

\begin{figure}
\epsscale{1.0}
\plotone{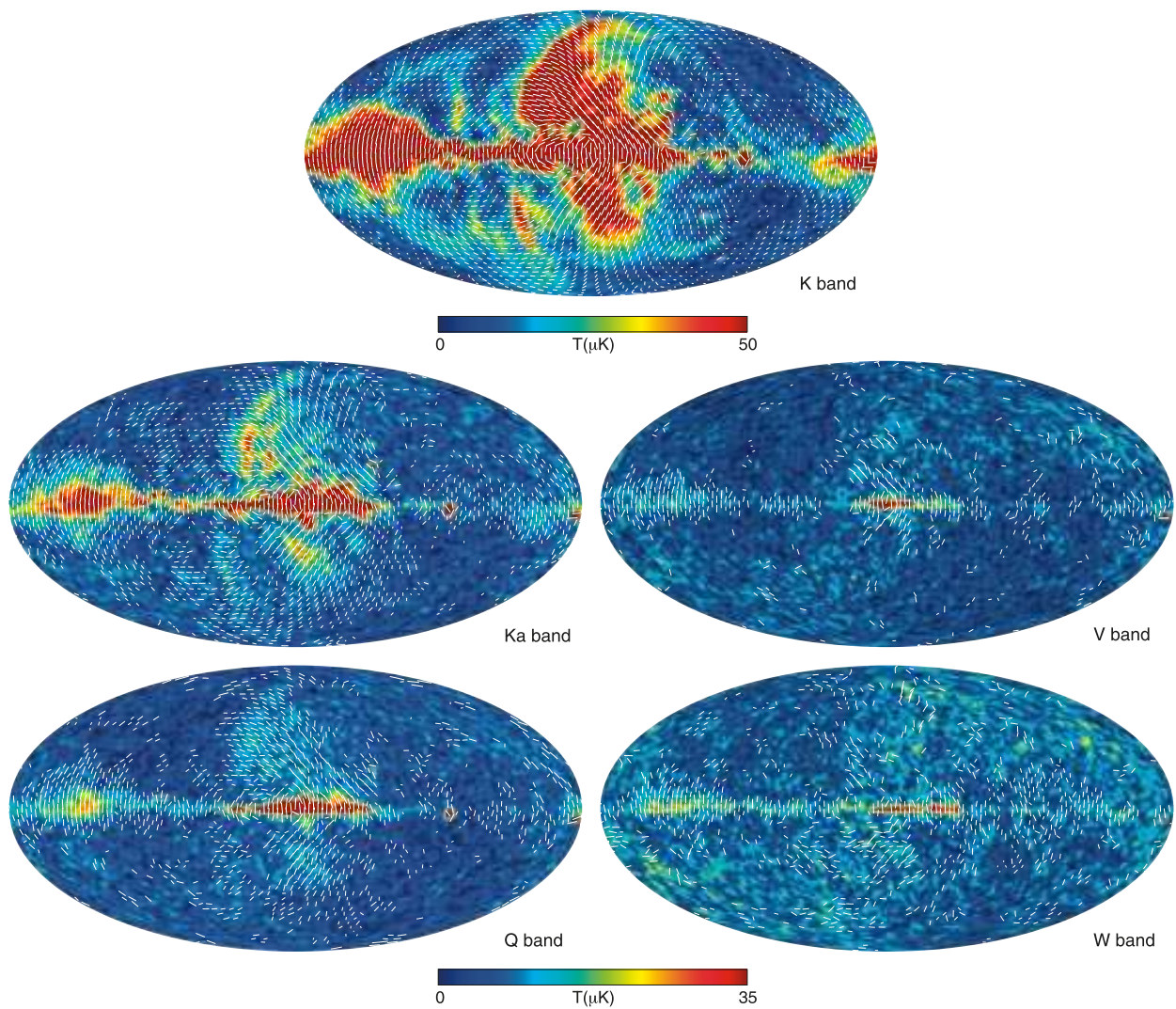}
\caption{Five-year polarization sky maps in Galactic coordinates smoothed to an
effective Gaussian beam of $2.0\dg$, shown in Mollweide projection. The color
scale indicates polarized intensity, $P = \sqrt{Q^2+U^2}$, and the line segments
indicate polarization direction in pixels whose signal-to-noise exceeds 1. {\it
top}: K band (23 GHz), {\it middle-left}: Ka band (33 GHz), {\it bottom-left}: Q
band (41 GHz), {\it middle-right}: V band (61 GHz), {\it bottom-right}: W band
(94 GHz).}
\label{fig:p_maps}
\end{figure}

After performing an end-to-end analysis of the instrument calibration,
single-year sky maps are created from the time-ordered data using the procedure
described by \citet{jarosik/etal:2007}.  Figure~\ref{fig:i_maps} shows the 5
year temperature maps at each of the five \map\ observing frequencies: 23, 33,
41, 61, and 94 GHz.  The number of independent observations per pixel, $N_{\rm
obs}$, is qualitatively the same as Figure~2 of \citet{hinshaw/etal:2007} and is
not reproduced here.  The noise per pixel, $p$, is given by $\sigma(p) =
\sigma_0 N_{\rm obs}^{-1/2}(p)$, where $\sigma_0$ is the noise per observation,
given in Table~\ref{tab:radiometers}.  To a very good approximation, the noise
per pixel in the 5 year maps is a factor of $\sqrt{5}$ times lower than in the
single-year maps.  Figures~\ref{fig:q_maps} and \ref{fig:u_maps} show the 5 year
polarization maps in the form of the Stokes parameters Q and U, respectively. 
Maps of the relative polarization sensitivity, the Q and U analogs of $N_{\rm
obs}$, are shown in Figure~13 of \citet{jarosik/etal:2007} and are not updated
here.  A description of the low-resolution pixel-pixel inverse covariance
matrices used in the polarization analysis is also given in
\citet{jarosik/etal:2007}, and is not repeated here.  The polarization maps are
dominated by foreground emission, primarily synchrotron emission from the Milky
Way.  Figure~\ref{fig:p_maps} shows the polarization maps in a form in which the
color scale represents polarized intensity, $P = \sqrt{Q^2+U^2}$, and the line
segments indicate polarization direction for pixels with a signal-to-noise ratio
greater than 1.  As with the temperature maps, the noise per pixel in the 5 year
polarization maps is $\sqrt{5}$ times lower than in the single-year maps.

\begin{figure}
\epsscale{0.9}
\plotone{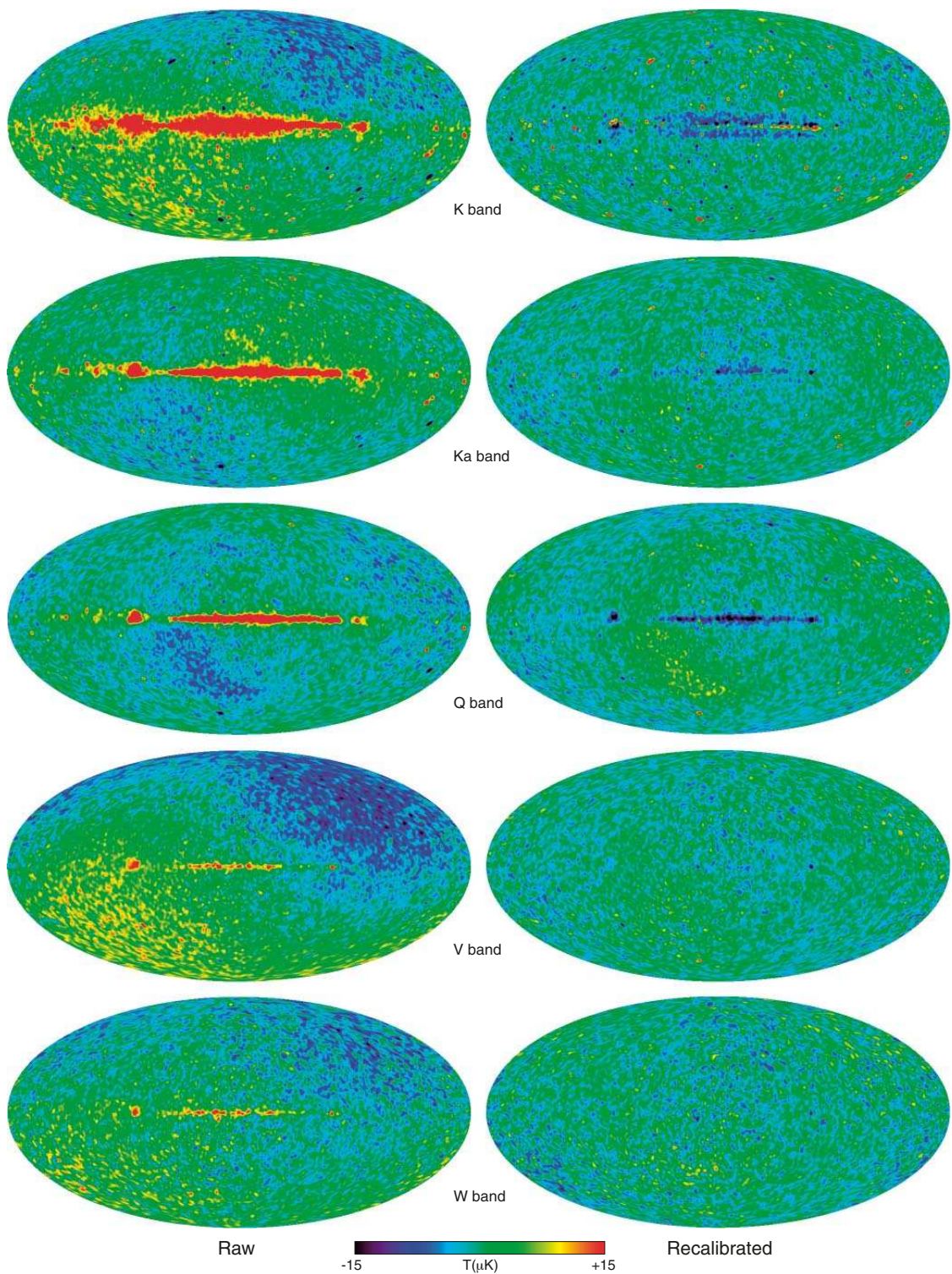}
\caption{Difference between the 5 year and 3 year temperature maps.  {\it left
column}: the difference in the maps, as delivered, save for the subtraction of a
relative offset (Table~\ref{tab:i_diff}), {\it right column}: the difference
after correcting the 3 year maps by a scale factor that accounts for the mean
gain change, $\sim 0.3$\%, between the 3 year and 5 year estimates.  {\it top to
bottom}: K, Ka, Q, V, W band.  The differences before recalibration are
dominated by galactic plane emission and a dipole residual: see
Table~\ref{tab:i_diff}, which also gives the changes for $l=2,3$.}
\label{fig:diff_maps_i_recal}
\end{figure}

Figure~\ref{fig:diff_maps_i_recal} shows the difference between the 5 year
temperature maps and the corresponding 3 year maps.  All maps have been smoothed
to 2$^{\circ}$ resolution to minimize the noise difference between them (due to
the additional years of data).  The left column shows the difference without any
further processing, save for the subtraction of a relative offset between the
maps.  Table~\ref{tab:i_diff} gives the value of the relative offset in each
band.  Recall that \map\ is insensitive to absolute temperature, so we adopt a
convention that sets the zero level in each map based on a model of the
foreground emission at the galactic poles.  While we have not changed
conventions, our 3 year estimate was erroneous due to the use of a preliminary
CMB signal map at the time the estimate was made.  This error did not affect any
cosmological results, but it probably explains the offset differences noted by
\citet{eriksen/etal:prep} in their recent analysis of the 3 year data.

The dominant structure in the left column of Figure~\ref{fig:diff_maps_i_recal}
consists of a residual dipole and galactic plane emission.  This reflects the
updated 5 year calibration which has produced changes in the gain of order 0.3\%
compared to the 3 year gain estimate (see \S\ref{sec:cal_improve} for a more
detailed discussion of the calibration).  Table~\ref{tab:i_diff} gives the
dipole amplitude difference in each band, along with the much smaller quadrupole
and octupole power difference.  (For comparison, we estimate the CMB power at
$l=2,3$ to be $l(l+1)C_l / 2\pi = 211, 1041$ $\mu$K$^2$, respectively.)  The
right column of Figure~\ref{fig:diff_maps_i_recal} shows the corresponding sky
map differences after the 3 year map has been rescaled by a single factor (in
each band) to account for the mean gain change between the 3 and 5 year
calibration determinations.  The residual galactic plane structure in these maps
is less than 0.2\% of the nominal signal in Q band, and less than 0.1\% in all
the other bands.  The large scale structure in the band-averaged temperature
maps is quite robust.

\subsection{CMB Dipole}
\label{sec:dipole}

The dipole anisotropy stands apart from the rest of the CMB signal due to its
large amplitude and to the understanding that it arises from our peculiar motion
with respect to the CMB rest frame.  In this section we present CMB dipole
results based on a new analysis of the 5 year sky maps.  Aside from an absolute
calibration uncertainty of 0.2\% (see \S\ref{sec:cal_improve}), the dominint
source of uncertainty in the dipole estimate arises from uncertainties in
Galactic foreground subtraction.  Here we present results for two different
removal methods: template-based cleaning and an internal linear combination
(ILC) of the \map\ multifrequency data \citep{gold/etal:prep}.  Our final
results are based on a combination of these methods with uncertainties that
encompass both approaches.

With template-based foreground removal, we can form cleaned maps for each of the
8 high frequency DA's, Q1-W4, while the ILC method produces one cleaned map from
a linear combination of all the \map\ frequency bands.  We analyze the residual
dipole moment in each of these maps (a nominal dipole based on the 3 year data
is subtracted from the time-ordered data prior to map-making) using a Gibbs
sampling technique which generates an ensemble of full-sky CMB realizations that
are consistent with the data, as detailed below.  We evaluate the dipole moment
of each full-sky realization and compute uncertainties from the scatter of the
realizations.

We prepared the data for the Gibbs analysis as follows.  The $N_{\rm side}=512$,
template-cleaned maps were zeroed within the KQ85 mask, smoothed with a
$10^{\circ}$ FWHM Gaussian kernel, and degraded to $N_{\rm side}=16$.  Zeroing
the masked region prior to smoothing prevents residual cleaning errors within
the mask from contaminating the unmasked data.  We add random white noise (12
$\mu$K rms per pixel) to each map to regularize the pixel-pixel covariance
matrix.  The $N_{\rm side}=512$ ILC map was also smoothed with a $10^{\circ}$
FWHM Gaussian kernel and degraded to $N_{\rm side}=16$, but the data within the
sky mask were not zeroed prior to smoothing.  We add white noise of 6 $\mu$K per
pixel to the smoothed ILC map to regularize its covariance matrix.  Note that
smoothing the data with a $10^{\circ}$ kernel reduces the residual dipole in the
maps by $\sim$0.5\%.  We ignore this effect since the residual dipole is only
$\sim$0.3\% of the full dipole amplitude to start with.

The Gibbs sampler was run for 10,000 steps for each of the 8 template-cleaned
maps (Q1-W4) and for each of 6 independent noise realizations added to the ILC
map.  In both cases we applied the KQ85 mask to the analysis and truncated the
CMB power at $l_{\rm max}=32$.  The resulting ensembles of 80,000 and 60,000
dipole samples were analyzed independently and jointly.  The results of this
analysis are given in Table~\ref{tab:dipole}.  The first row combines the
results from the template-cleaned DA maps; the scatter among the 8 DA's was well
within the noise scatter for each DA, so the Gibbs samples for all 8 DA's were
combined for this analysis.  The results for the ILC map are shown in the second
row.  The two methods give reasonably consistent results, however, the Galactic
longitude of the two dipole axis estimates differ from each other by about
2$\sigma$.  Since we cannot reliably identify one cleaning method to be superior
to the other, we have merged the Gibbs samples from both methods to produce the
conservative estimate shown in the bottom row.  This approach, which enlarges
the uncertainty to emcompass both estimates, gives
\beq
(d,l,b) = (3.355 \pm 0.008 \; {\rm mK}, 263.99^{\circ} \pm 0.14^{\circ}, 48.26^{\circ} \pm 0.03^{\circ}),
\eeq
where the amplitude estimate includes the 0.2\% absolute calibration
uncertainty.  Given the CMB monopole temperature of 2.725 K
\citep{mather/etal:1999}, this amplitude implies a Solar System peculiar
velocity of $369.0 \pm 0.9$ km s$^{-1}$ with respect to the CMB rest frame.

\section{CALIBRATION IMPROVEMENTS}
\label{sec:cal_improve}

With the 5 year processing we have refined our procedure for evaluating the
instrument calibration, and have improved our estimates for the calibration
uncertainty.  The fundamental calibration source is still the dipole anisotropy
induced by \map's motion with respect to the CMB rest frame
\citep{hinshaw/etal:2003b, jarosik/etal:2007}, but several details of the
calibration fitting have been modified.  The new calibration solution is
consistent with previous results in the overlapping time range.  We estimate the
uncertainty in the absolute calibration is now 0.2\% per differencing assembly.

The basic calibration procedure posits that a single channel of time-ordered
data, $d_i$, may be modeled as
\beq
d_i = g_i \left[\Delta T_{vi} + \Delta T_{ai} \right] + b_i,
\eeq
where $i$ is a time index, $g_i$ and $b_i$ are the instrument gain and baseline,
at time step $i$, $\Delta T_{vi}$ is the differential dipole anisotropy induced
by \map's motion, and $\Delta T_{ai}$ is the differential sky anisotropy.  We
assume that $\Delta T_{vi}$ is known exactly and has the form
\beq
\Delta T_{vi} = \frac{T_0}{c} {\bf v}_i \cdot [(1+x_{\rm im}){\bf n}_{A,i} - (1-x_{\rm im}){\bf n}_{B,i}],
\label{eq:Tv}
\eeq
where $T_0 = 2.725$ K is the CMB temperature \citep{mather/etal:1999}, $c$ is
the speed of light, ${\bf v}_i$ is \map's velocity with respect to the CMB rest
frame at time step $i$, $x_{\rm im}$ is the loss imbalance parameter
\citep{jarosik/etal:2007}, and ${\bf n}_{A,i}$, and ${\bf n}_{B,i}$ are the unit
vectors of the A- and B-side lines of sight at time step $i$ (in the same frame
as the velocity vector).  The velocity may be decomposed as
\beq
{\bf v}_i = {\bf v}_{\rm WMAP-SSB,i} + {\bf v}_{\rm SSB-CMB},
\eeq
where the first term is \map's velocity with respect to the solar system
barycenter, and the second is the barycenter velocity with respect to the CMB. 
The former is well determined from ephemeris data, while the latter has been
measured by COBE-DMR with an uncertainty of 0.7\% \citep{Kogut/etal:1996d}. 
Since the latter velocity is constant over \map's life span, any error in our
assumed value of ${\bf v}_{\rm SSB-CMB}$ will, in theory, be absorbed into a
dipole contribution to the anisotropy map, $T_a$.  We test this hypothesis
below.  The differential sky signal has the form
\beq
\Delta T_{ai} = (1+x_{\rm im})[I_a(p_{A,i})+P_a(p_{A,i},\gamma_{A,i})]
              - (1-x_{\rm im})[I_a(p_{B,i})+P_a(p_{B,i},\gamma_{B,i})],
\eeq
where $p_{A,i}$ is the pixel observed by the A-side at time step $i$ (and
similarly for B), $I_a(p)$ is the temperature anisotropy in pixel $p$ (the
intensity Stokes parameter, $I$), and $P_a(p,\gamma)$ is the polarization
anisotropy in pixel $p$ at polarization angle $\gamma$
\citep{hinshaw/etal:2003b} which is related to the linear Stokes parameters $Q$
and $U$ by
\beq
P_a(p,\gamma) = Q(p) \cos 2\gamma + U(p) \sin 2\gamma.
\eeq
We further note that, in general, $I_a$ and $P_a$ depend on frequency owing to
Galactic emission.

A main goal of the data processing is to simultaneously fit for the calibration
and sky signal.  Unfortunately, since the data model is nonlinear and the number
of parameters is large, the general problem is intractable.  In practice, we
proceed iteratively as follows.  Initially we assume the gain and baseline are
constant for a given time interval, typically between 1 and 24 hours, \beqa g_i
= G_k  &   & \tau_k < t_i < \tau_{k+1} \\ b_i = B_k  &   & \tau_k < t_i <
\tau_{k+1}, \eeqa where $t_i$ is the time of the $i$th individual observation,
and $\tau_k$ is the start time of the $k$th calibration interval.  Throughout
the fit we fix the velocity-induced signal, equation~(\ref{eq:Tv}), using ${\bf
v}_{\rm SSB-CMB} = [-26.29, -244.96, +275.93]$ km s$^{-1}$ (in Galactic
coordinates), and, for the first iteration, we assume no anisotropy signal,
$\Delta T_a = 0$.  Then, for each calibration interval $k$ we perform a linear
fit for $G_k$ and $B_k$ with fixed $\Delta T_v + \Delta T_a$.  As we proceed
through the intervals, we apply this calibration to the raw data and accumulate
a new estimate of the anisotropy map as per equation~19 of
\citet{hinshaw/etal:2003b}.  The procedure is repeated with each updated
estimate of $\Delta T_a$.  Once the calibration solution has converged, we fit
the gain data, $G_k$, to a model that is parameterized by the instrument
detector voltage and the temperatures of the receiver's warm and cold stages,
equation~2 of \citet{jarosik/etal:2007}.  This parametrization still provides a
good fit to the $G_k$ data, so we have not updated its form for the 5 year
analysis.  The updated best-fit parameters are given in the 5 year Explanatory
Supplement \citep{limon/etal:prep}.  Note that for each radiometer, the relative
gain vs. time over 5 years is determined by just two parameters.

For the 5 year processing we have focused on the veracity of the ``raw''
calibration, $G_k$ and $B_k$.  Specifically, we have improved and/or critically
reexamined several aspects of the iterative fitting procedure:
\begin{itemize}
\item We have incorporated the effect of far sidelobe pickup directly into the
iterative calibration procedure, rather than as a fixed correction
\citep{jarosik/etal:2007}.   We do this by segregating the differential signal
into a main beam contribution and a sidelobe contribution,
\beq
\Delta T_i = \Delta T_{\rm main,i} + \Delta T_{\rm side,i}.
\eeq
(\citealt{hill/etal:prep} discuss how this segregation is defined in the 5 year
processing.)  After each iteration of the calibration and sky map estimation, we
(re)compute a database of $\Delta T_{\rm side}$ on a grid of pointings using the
new estimate of $I_a$.  We then interpolate the database to estimate $\Delta
T_{\rm side,i}$ for each time step $i$.  Note that $\Delta T_{\rm side}$
includes contributions from both the velocity-induced signal and the intrinsic
anisotropy.  Ignoring sidelobe pickup can induce gain errors of up to 1.5\% in K
band, 0.4\% in Ka band, and $\sim$0.25\% in Q-W bands.

\item In general, the different channels within a DA have different center
frequencies \citep{jarosik/etal:2003}; hence the different channels measure a
slightly different anisotropy signal due to differences in the Galactic signal. 
We assess the importance of accounting for this in the calibration procedure.

\item A single DA channel is only sensitive to a single linear polarization
state.  (\map~measures polarization by differencing orthogonal polarization
channels.)  Thus we cannot reliably solve for both $P_a$ and for $I_a$ at each
channel's center frequency.  We assess the relative importance of accounting for
one or the other on both the gain and baseline solutions.

\item We examine the sensitivity of the calibration solution to the choice of
${\bf v}_{\rm SSB-CMB}$ and to assumptions of time-dependence in the gain.

\end{itemize}

\subsection{Calibration Tests}
\label{sec:cal_test}

We use a variety of end-to-end simulations to assess and control the systematic
effects noted above.  We summarize a number of the key tests in the remainder of
this section.

\begin{figure}
\epsscale{0.75}
\plotone{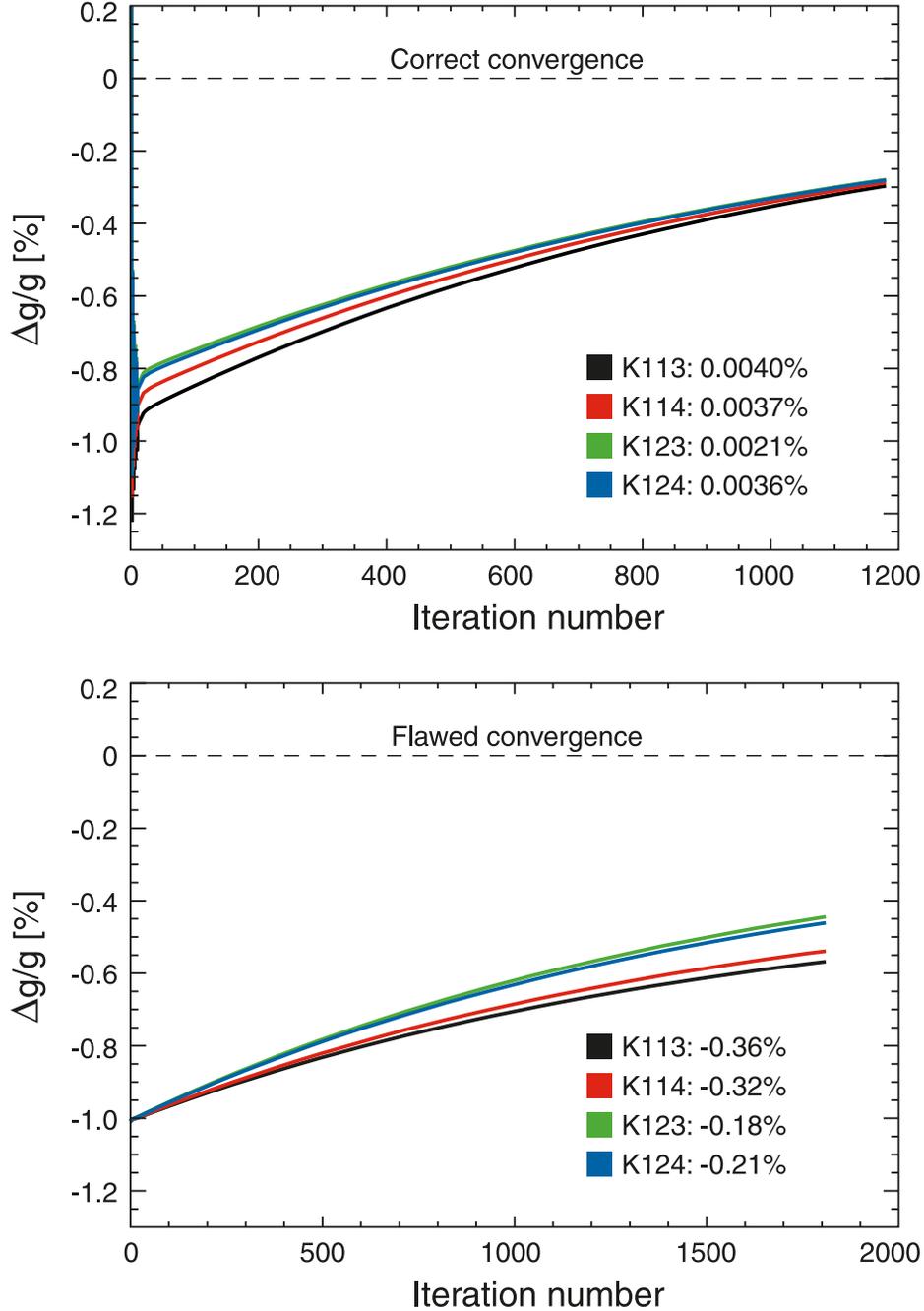}
\caption{Gain convergence tests using the iterative sky map \& calibration
solver run on a pair of simulations with known, but different, inputs.  Both
panels show the recovered gain as a function of iteration number for a 4-channel
K band simulation.  The initial calibration guess was chosen to be in error by
1\% to test convergence; the output solutions, extrapolated with an exponential
fit, are printed in each panel.  {\it top}:  Results for a noiseless simulation
that includes a dipole signal (with Earth-velocity modulation) plus CMB and
foreground anisotropy (the former is evaluated at the center frequency of each
channel).  The input gain was set to be constant in time.  The extrapolated
solutions agree with the input values to much better than 0.1\%.  {\it bottom}: 
Results for a noiseless simulation that includes only dipole signal (with
Earth-velocity modulation) but no CMB or foreground signal.  In this case the
input gain was set up to have flight-like thermal variations. The extrapolated
absolute gain recovery was in error by $\gt$0.3\%, indicating a small residual
degeneracy between the sky model and the time-dependent calibration.}
\label{fig:gain_converge}
\end{figure}

The first case we consider is a noiseless simulation in which we generate
time-ordered data from an input anisotropy map which includes CMB and Galactic
foreground signal (one map per channel, evaluated at the center frequency of
each channel) and a known dipole amplitude.  The input gain for each channel is
fixed to be constant in time. We run the iterative calibration and sky map
solver allowing for an independent sky map solution at each channel (but no
polarization signal).  When fitting for the calibration, we assume that ${\bf
v}_{\rm SSB-CMB}$ differs from the input value by 1\% to see if the known,
modulated velocity term, ${\bf v}_{\rm WMAP-SSB}$, properly ``anchors'' the
absolute gain solution.  The results are shown in the top panel of
Figure~\ref{fig:gain_converge} where it is shown that the absolute gain recovery
is robust to errors in ${\bf v}_{\rm SSB-CMB}$.  We recover the input gain to
better than 0.1\% in this instance.

The second case we consider is again a noiseless simulation that now includes
{\it only} dipole signal (with Earth-velocity modulation), but here we vary the
input gain using the flight-derived gain model \citep{jarosik/etal:2007}.  The
iterative solver was run on the K band data for 1400 iterations, again starting
with an initial guess that was in error by 1\%.  The results are shown in the
bottom panel of Figure~\ref{fig:gain_converge}, which indicate systematic
convergence errors of $\gt$0.3\% in the fitted amplitude of the recovered gain
model.  Since the input sky signal in this case does not have any Galactic
foreground or polarization components, we cannot ascribe the recovery errors to
the improper handling of those effects in the iterative solver.  We have also
run numerous other simulations that included various combinations of instrument
noise, CMB anisotropy, Galactic foreground signal (with or without individual
center frequencies per channel), polarization signal, and input gain
variations.  The combination of runs are too numerous to report on in detail,
and the results are not especially enlightening.  The most pertinent trend we
can identify is that when the input value of  ${\bf v}_{\rm SSB-CMB}$ is assumed
in the iterative solver, the recovered gain is in good agreement with the input,
but when the initial guess is in error by 1\%, the recovered gain will have
comparable errors.  We believe the lack of convergence is due to a weak
degeneracy between gain variations and the sky map solution.  Such a degeneracy
is difficult to diagnose in the context of this iterative solver, especially
given the computational demands of the system, so we are assessing the
system more directly with a low-resolution parameterization of the gain and sky
signal, as outlined in Appendix~\ref{app:cal_fisher}.

\begin{figure}
\epsscale{1.0}
\plotone{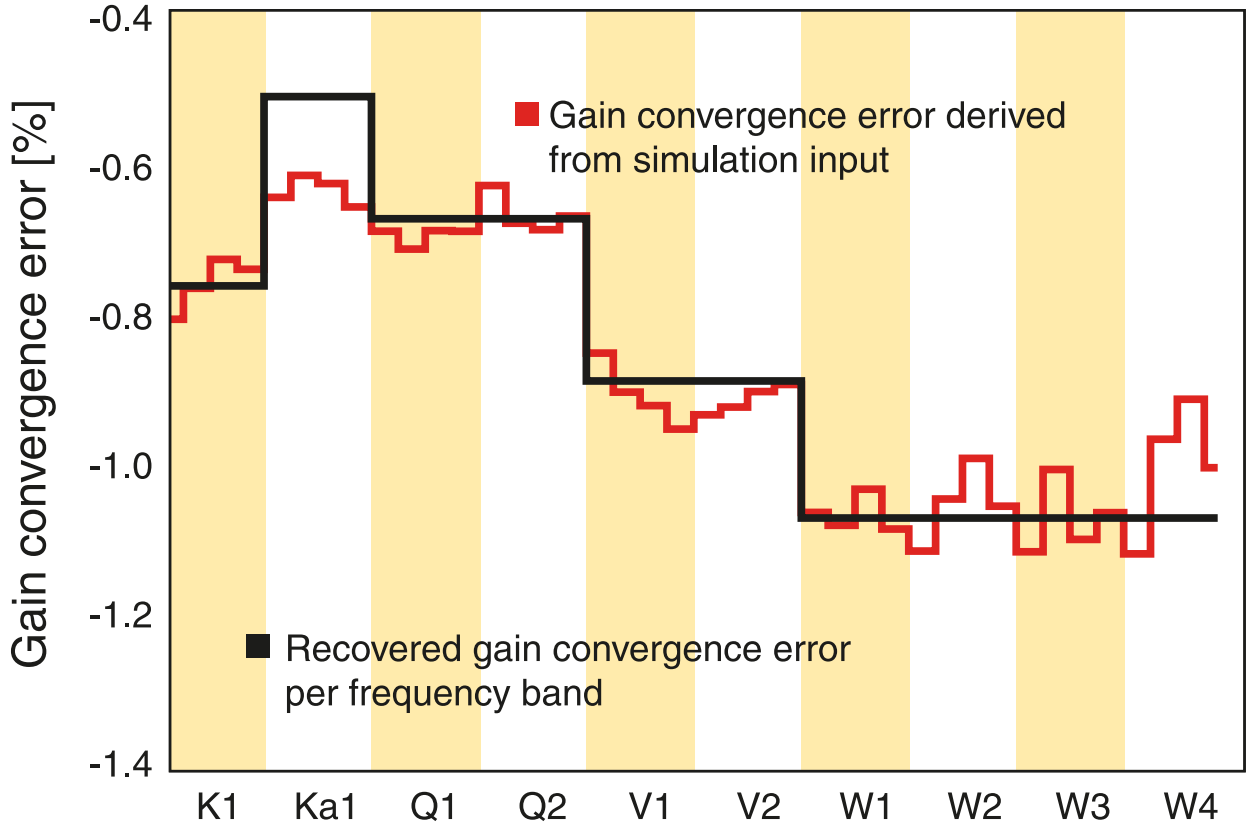}
\caption{Gain error recovery test from a flight-like simulation that includes
every effect known to be important.  Using the daily dipole gains recovered from
the iterative sky map \& calibration solver as input, the gain convergence
error, shown here, is fit simultaneously with the gain model parameters, not
shown, following the procedure outlined in Appendix~\ref{app:cal_model_fit}. 
The red trace indicates the true gain error for each \map\ channel, based on the
known input gain and the gain solution achieved by the iterative solver on its
final iteration.  The black trace shows the gain error recovered by the fit,
averaged by frequency band.  The channel-to-channel scatter within a band is
$\lt$0.1\%, though the mean of Ka band error is of order 0.1\%.}
\label{fig:gain_recover}
\end{figure}

Since the latter effort is still underway, we have adopted a more pragmatic
approach to evaluating the absolute gain and its uncertainty for the 5 year data
release.  We proceed as follows: after 50 iterations of the calibration and sky
map solver, the dominant errors in the gain and sky map solution are 1) a dipole
in the sky map, and 2) a characteristic wave form that reflects a relative error
between ${\bf v}_{\rm SSB-CMB}$ and ${\bf v}_{\rm WMAP-SSB}$.   At this point we
can calibrate the amplitude of the gain error wave form to the magnitude of the
velocity error in ${\bf v}_{\rm SSB-CMB}$.  We can then fit the gain solution to
a linear combination of the gain model of \citet{jarosik/etal:2007} and the
velocity error wave form.  See Appendix~\ref{app:cal_model_fit} for details on
this fitting procedure.  In practice this fit is performed simultaneously on
both channels of a radiometer since those channels share one gain model
parameter.  We have tested this procedure on a complete flight-like simulations
that includes every important effect known, including input gain variations. 
The results of the gain recovery are shown in Figure~\ref{fig:gain_recover}, and
based on this we conservatively assign an absolute calibration uncertainty of
0.2\% per channel for the 5 year \map\ archive.

\subsection{Summary}
\label{sec:cal_summ}

The series of steps taken to arrive at the final 5 year calibration are as
follows:
\begin{itemize}

\item Run the iterative calibration and sky map solver over the full 5 year data
set for 50 iterations, using 24 hour calibration intervals.  This run starts
with $I_a = P_a = 0$ and updates $I_a$ for each individual channel of data. 
$P_a$ is assumed to be 0 throughout this run.  We keep the gain solution, $G_k$,
from this run and discard the baseline solution.

\item Run the iterative calibration and sky map solver over the full 5 year data
set for 50 iterations, using 1 hour calibration intervals.  This run starts with
$I_a = P_a = 0$ and updates both using the intensity and polarization data in
the two radiometers per DA, as per Appendix~D of \citet{hinshaw/etal:2003b}.  We
keep the baseline solution, $B_k$, from this run and discard the gain solution. 
Both of these runs incorporate the sidelobe correction as noted above.

\item Fit the gain solution, $G_k$ simultaneously for the gain model and for an
error in the velocity, ${\bf \Delta v}_{\rm SSB-CMB}$, as described in
Appendix~\ref{app:cal_model_fit}.  This fit is performed on two channels per
radiometer with the gain model parameter $T_0$ common to both channels.

\item We average the best-fit velocity error over all channels within a
frequency band under the assumption that the dipole is the same in each of these
channels.  We then fix the velocity error to a single value per frequency band
and re-fit the gain model parameters for each pair of radiometer channels. 

\end{itemize}

Based on end-to-end simulations with flight-like noise, we estimate the absolute
gain error per radiometer to be 0.2\%.  We believe the limiting factor in this
estimate is a weak degeneracy between thermal variations in the instrument gain,
which are annually modulated, and annual variations induced by errors in ${\bf
v}_{\rm SSB-CMB}$.  Since there is a small monotonic increase in the spacecraft
temperature, additional years of data should allow improvements in our ability
to separate these effects.

Once we have finalized the gain model, we form a calibrated time-ordered data
archive using the gain model and the 1 hour baseline estimates to calibrate the
data.  This archive also has a final estimate of the far sidelobe pickup
subtracted from each time-ordered data point.  However, we opt not to subtract a
dipole estimate from the archive at this stage in the processing.

\section{BEAM IMPROVEMENTS}
\label{sec:beams}

In addition to reassessing the calibration, the other major effort undertaken to
improve the 5 year data processing was to extend the physical optics model of
the \map\ telescope based on flight measurements of Jupiter.  This work is
described in detail in \citet{hill/etal:prep} so we only summarize the key
results with an emphasis on their scientific implications.  The basic aim of the
work is to use the flight beam maps from all 10 DA's to determine the in-flight
distortion of the mirrors.  This program was begun for the A-side mirror during
the 3 year analysis; for the 5 year analysis we have quadrupled the number of
distortion modes we fit (probing distortion scales that are half the previous
size), and we have developed a completely new and independent model of the
B-side distortions, rather than assuming that they mirror the A-side
distortions.  We have also placed limits on smaller scale distortions by
comparing the predicted beam response at large angles to sidelobe data collected
during \map's early observations of the Moon.

Given the best-fit mirror model, we compute the model beam response for each DA
and use it in conjunction with the flight data to constrain the faint tails of
the beams, beyond $\sim 1^{\circ}$ from the beam peak.  These tails are
difficult to constrain with flight data alone because the Jupiter signal to
noise ratio is low, but, due to their large areal extent they contain a
non-negligible fraction (up to 1\%) of the total beam solid angle.  An accurate
determination of the beam tail is required to properly measure the ratio of
sub-degree-scale power to larger-scale power in the diffuse CMB emission (and to
accurately assign point source flux).

Figure~14 in \citet{hill/etal:prep} compares the beam radial profiles used in
the 3 year and 5 year analyses, while Figure~13 compares the $l$-space transfer
functions derived from the Legendre transform of the radial profile.  The
important changes to note are the following.

\begin{enumerate}

\item In both analyses we split the beam response into main beam and far
sidelobe contributions.  In the 5 year analysis we have enlarged the radius at
which this transition is made \citep{hill/etal:prep}.  In both cases, we correct
the time-ordered data for far sidelobe pickup prior to making sky maps, while
the main beam contribution is only accounted for in the analysis of sky maps,
e.g., in power spectrum deconvolution.  As a result, the sky maps have a
slightly different effective resolution which is most apparent in K band, as in
Figure~\ref{fig:diff_maps_i_recal}.  However, in each analysis, the derived
transfer functions are appropriate for the corresponding sky maps.

\item In the 3 year analysis, the main beam profile was described by a Hermite
polynomial expansion fit to the observations of Jupiter in the time-ordered
data.  This approach was numerically problematic in the 5 year analysis due to
the larger transition radius; as a result, we now simply co-add the time-ordered
data into radial bins to obtain the profiles.  In both cases, the underlying
time-ordered data is a hybrid archive consisting of flight data for points where
the beam model predicts a value above a given contour, and model values for
points below the contour \citep{hill/etal:prep}.  With the improved beam models
and a new error analysis, we have adjusted these hybrid contours down slightly,
with the result that we use proportionately more flight data (per year) in the
new analysis.  The radius at which the 5 year profile becomes model dominated
($\gt$50\% of the points in a bin) is indicated by dotted lines in Figure~14 of
\citet{hill/etal:prep}.

\item The right column of Figure~14 in \citet{hill/etal:prep} shows the
fractional change in solid angle due to the updated profiles.  The main point to
note is the $\sim$1\% increase in the V2 and W band channels, primarily arising
in the bin from 1 to 2 degrees off the beam peak.  As can be seen in Figure~3 of
\citet{hill/etal:prep}, this is the angular range in which the new beam models
produced the most change, owing to the incorporation of smaller distortion modes
in the mirror model.  The 3 year analysis made use of the model in this angular
range which, in hindsight, was suppressing up to $\sim$1\% of the solid angle in
the V and W band beams.  (The longer wavelength channels are less sensitive to
distortions in this range, so the change in solid angle is smaller for K-Q
bands.)  In the 5 year analysis, we use relatively more flight data in this
regime, so we are less sensitive to any remaining model uncertainties. 
\citet{hill/etal:prep} place limits on residual model errors and propagate those
errors into the overall beam uncertainty.

\item Figure~13 in \citet{hill/etal:prep} compares the beam transfer functions,
$b_l$, derived by transforming the 3 year and 5 year radial profiles.  (To
factor out the effect of changing the transition radius, the 3 year profiles
were extended to the 5 year radius using the far sidelobe data, for this
comparison.)  Since the transfer functions are normalized to 1 at $l=1$, the
change is restricted to high $l$.  In V and W bands, $b_l$ has decreased by
$\sim$0.5 - 1\% due largely to the additional solid angle picked up in the 1-2
degree range.  This amounts to a $\sim$1 $\sigma$ change in the functions, as
indicated by the red curves in the Figure.

\end{enumerate}

The calibrated angular power spectrum is proportional to $1/g^2 b_l^2$, where
$g$ is the mean gain and $b_l$ is the beam transfer function, thus the net
effect of the change in gain and beam determinations is to increase the power
spectrum by $\sim$0.5\% at $l \la 100$, and by $\sim$2.5\% at high $l$. 
\citet{nolta/etal:prep} give a detailed evaluation of the power spectrum while
\citet{dunkley/etal:prep} and \citet{komatsu/etal:prep} discuss the implications
for cosmology.

\section{LOW-$l$ POLARIZATION TESTS}
\label{sec:pol_test}

The 3 year data release included the first measurement of microwave polarization
over the full sky, in the form of Stokes Q and U maps in each of 5 bands.  The
analysis of \map\ polarization data is complicated by the fact that the
instrument was not designed to be a true polarimeter, thus a number of
systematic effects had to be understood prior to assigning reliable error
estimates to the data.  \citet{page/etal:2007} presented the 3 year polarization
data in great detail.  In this section we extend that analysis by considering
some additional tests that were not covered in the 3 year analysis. We note that
all of the tests described in this section have been performed on the
template-cleaned reduced-foreground maps except for the final test of the Ka
band data, described at the end of the section, which tests an alternative
cleaning method.

\subsection{Year-to-Year Consistency Tests}
\label{sec:pol_year}

With 5 years of data it is now possible to subject the data to more stringent
consistency tests than was previously possible.  In general, the number of
independent cross-power spectra we can form within a band with $N_d$
differencing assemblies is $N_d(N_d-1)/2 \times N_y + N_d^2 \times
N_y(N_y-1)/2$.  With 5 years of data, this gives 10 independent estimates each
in K and Ka band, 45 each in Q and V band, and 190 in W band.  For cross power
spectra of distinct band pairs, with $N_{d1}$ and $N_{d2}$ DA's in each band,
the number is $N_{d1}N_{d2} \times N_y^2$.  This gives 50 each in KaQ and KaV,
100 each in KaW and QV, and 200 each in QW and VW.  (For comparison, the
corresponding numbers are 3, 15, \& 66, and 18, 36, \& 72 with 3 years of
data.)

We have evaluated these individual spectra from the 5 year data and have
assigned noise uncertainties to each estimate using the Fisher formalism
described in \citet{page/etal:2007}.  We subject the ensemble to an internal
consistency test by computing the reduced $\chi^2$ of the data at each multipole
$l$ within each band or band pair, under the hypothesis that the data at each
multipole and band measures the same number from DA to DA and year to year.  The
results of this test are given in Table~\ref{tab:chi2_pol_l} for the
foreground-cleaned EE, EB, and BB spectra from $l=2-10$ for all band pairs from
KaKa to WW.  There are several points to note in these results.

\begin{enumerate} 

\item For $l \ge 6$, the most significant deviation from 1 in reduced $\chi^2$,
in any spectrum or band, is 1.594 in the $l=7$ BB spectrum for KaQ.  With 50
degrees of freedom, this is a 3 $\sigma$ deviation, but given that we have 150
$l \ge 6$ samples in the table, we expect of order 1 such value.  Thus we
conclude that the Fisher-based errors provide a good description of the DA-to-DA
and year-to-year scatter in the $l \ge 6$ polarization data.  If anything, there
is a slight tendency to overestimate the uncertainties at higher $l$.

\item For $l \le 5$, we find 37 out of 120 points where the reduced $\chi^2$
deviates from 1 at more than 4 $\sigma$ significance, indicating excessive
internal scatter in the data relative to the Fisher errors.  However, all but 5
of these occur in cross-power spectra in which one or both of the bands contain
W band data.  If we exclude combinations with W band, the remaining 72 points
have a mode in the reduced $\chi^2$ distribution of 1 with a slight positive
skewness due to the 5 points noted above, which all contain Q band data.  This
may be a sign of slight foreground residuals contributing additional noise to
the Q band data, though we do not see similar evidence in the Ka band spectra
which would be more foreground contaminated prior to cleaning.  For Ka-V bands,
we believe that the Fisher errors provide an adequate description of the scatter
in this $l \le 5$ polarization data, but we subject polarization sensitive
cosmological parameter estimates, e.g., the optical depth, to additional
scrutiny in \S\ref{sec:pol_Ka}.

\item Of special note is $l=3$ BB which, as noted in \citet{page/etal:2007}, is
the power spectrum mode that is least modulated in the \map\ time-ordered data. 
This mode is therefore quite sensitive to how the instrument baseline is
estimated and removed and, in turn, to how the 1/f noise is modeled.  In the
accounting above, the $l=3$ BB data have the highest internal scatter of any
low-$l$ polarization mode.  In particular, {\em every} combination that includes
W band data is significantly discrepant; and the two most discrepant non-W band
points are also estimates of $l=3$ BB.  We comment on the W band data further
below, but note here that the final co-added BB spectrum (based on Ka, Q, and V
band data) does {\em not} lead to a significant detection of tensor modes. 
However, we caution that any surprising scientific conclusions which rely
heavily on the \map\ $l=3$ BB data should be treated with caution.

\end{enumerate}

Based on the analysis presented above,  we find the W band polarization data is
still too unstable at low-$l$ to be reliably used for cosmological studies.  We
cite more specific phenomenology and consider some possible explanations in the
remainder of this section.

\begin{figure}
\epsscale{1.0}
\plotone{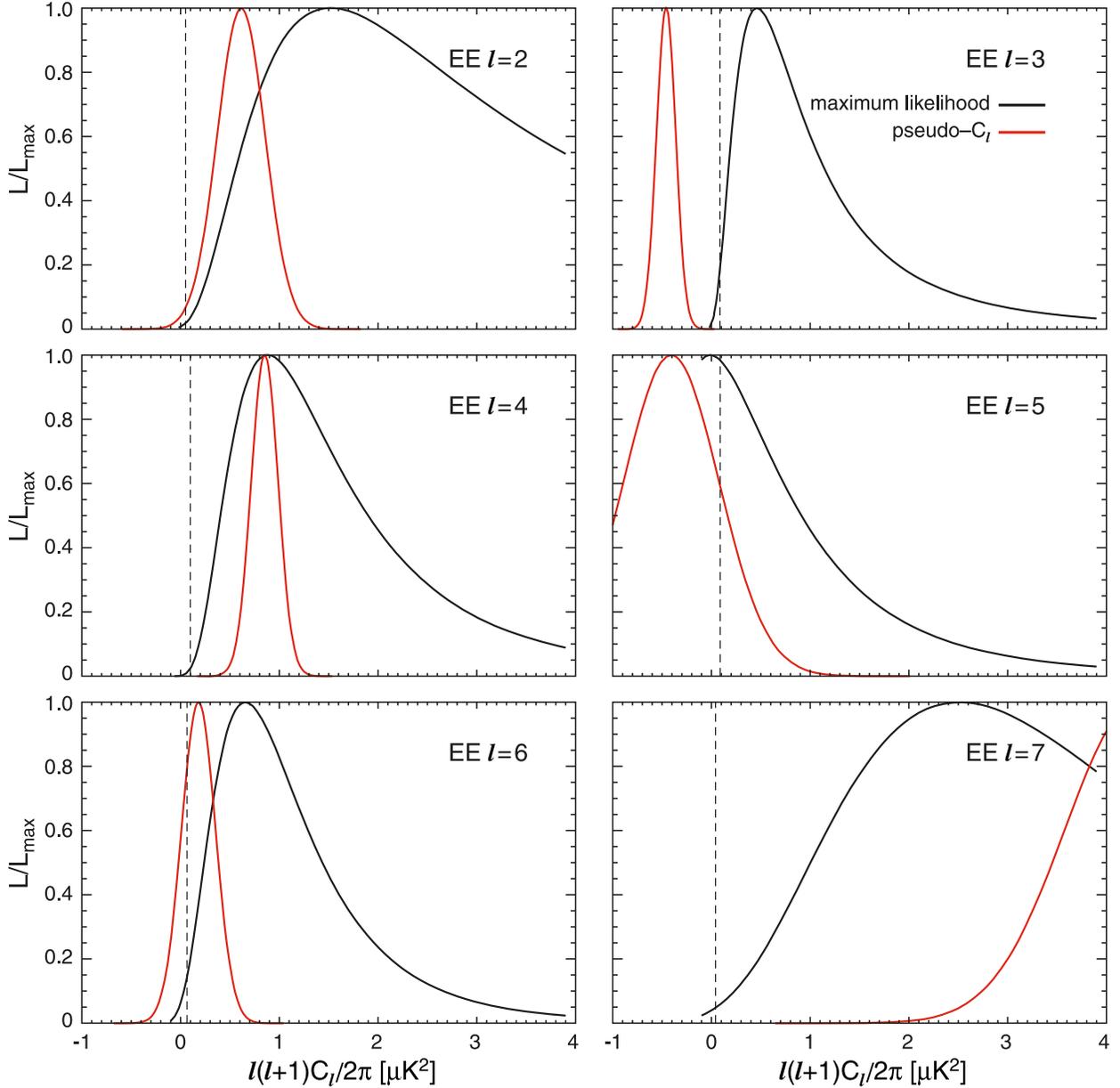}
\caption{W band EE power spectrum likelihood from $l=2-7$ using two separate
estimation methods: {\it black}: maximum likelihood and {\it red}:
pseudo-$C_l$.  The vertical dashed lines indicate the best-fit model power
spectrum based on fitting the combined Ka, Q, and V band data.  The two spectrum
estimates are consistent with each other, except at $l=3$.  The maximum
likelihood estimates are wider because they include cosmic variance whereas the
pseudo-$C_l$ estimates account for noise only.  Both estimates show excess power
in the W band data relative to the best-fit model, and to the combined KaQV band
spectrum, shown in Figure~6 of \citet{nolta/etal:prep}.  The extreme excess in
the $l=7$ pseudo-$C_l$ estimate is not so severe in the maximum likelihood, but
both methods are still inconsistent with the best-fit model.}
\label{fig:ww_ee}
\end{figure}

The 5 year co-added W band EE spectrum is shown in Figures~\ref{fig:ww_ee}, in
the form of likelihood profiles from $l=2-7$.  At each multipole we show two
curves: an estimate based on evaluating the likelihood multipole by multipole,
and an estimate based on the pseudo-$C_l$ method \citep{page/etal:2007}.  The
best-fit model EE spectrum, based on the combined Ka, Q, and V band data is
indicated by the dashed lines in each panel.  Both spectrum estimates show
excess power relative to the model spectrum, with the most puzzling multipole
being $l=7$ which, as shown in Table~\ref{tab:chi2_pol_l}, has an internal
reduced $\chi^2$ of 1.015, for 190 degrees of freedom.  This data has the
hallmark of a sky signal, but that hypothesis is implausible for a variety of
reasons \citep{page/etal:2007}.  It is more likely due to a systematic effect
that is common to a majority of the W band channels over a majority of the 5
years of data.  We explore and rule out one previously neglected effect in
\S\ref{sec:pol_emiss}.  It is worth recalling that $l=7$ EE, like $l=3$ BB, is a
mode that is relatively poorly measured by \map, as discussed in
\citet{page/etal:2007}; see especially Figure~16 and its related discussion.

The W band BB data also exhibit unusual behavior at $l=2,3$.  In this case,
these two multipoles have internal reduced $\chi^2$ greater than 6, and the
co-added $l=2$ point is nearly 10 $\sigma$ from zero.  However, with 190 points
in each 5 year co-added estimate it is now possible to look for trends within
the data that were relatively obscure with only 3 years of data.  In particular,
we note that in the $l=2$ estimate, there are 28 points that are individually
more than 5 $\sigma$ from zero and that {\em all} of them contain W1 data in one
or both of the DA pairs in the cross power spectrum.  Similarly for $l=3$, there
are 14 points greater than 5 $\sigma$ and {\em all} of those points contain W4
data in one or both of the DA pairs.  We have yet to pinpoint the significance
of this result, but we plan to study the noise properties of these DA's beyond
what has been reported to date, and to sharpen the phenomenology with additional
years of data.

\subsection{Emissivity Tests}
\label{sec:pol_emiss}

In this section we consider time dependent emission from the \map\ optics as a
candidate for explaining the excess W band ``signal'' seen in the EE spectrum,
mostly at $l=7$.  In the end, the effect proved not to be significant, but it
provides a useful illustration of a common-mode effect that we believe is still
present in the W band polarization data.

From a number of lines of reasoning, we know that the microwave emissivity of
the mirrors is a few percent in W band, and that it scales with frequency
roughly like $\nu^{1.5}$ across the \map\ frequency range, as expected for a
classical metal \citep{born/wolf:POO:6e}.  Hence this mechanism has the
potential to explain a common-mode effect that is primarily seen in W band. 
Further, Figure~1 in \citet{jarosik/etal:2007} shows that the physical
temperature of the primary mirrors are modulated at the spin period by $\sim$200
$\mu$K, with a dependence on solar azimuth angle that is highly repeatable from
year to year. We believe this modulation is driven by solar radiation
diffracting around the \map\ sun shield reaching the tops of the primary
mirrors, which are only a few degrees within the geometric shadow of the sun
shield.  In contrast, the secondary mirrors and feed horns are in deep shadow
and show no measurable variation at the spin period, so that any emission they
produce only contributes to an overall radiometer offset, and will not be
further considered here.  

As a rough estimate, the spin modulated emission from the primary mirrors could
produce as much as $\sim 0.02 \times 200 = 4$ $\mu$K of radiometric response in
W band, but the actual signal depends on the relative phase of the A and B-side
mirror variations and the polarization state of the emission.  In more detail,
the differential signal, $d(t)$, measured by a radiometer with lossy elements is
\beq
d(t) = (1 - \epsilon_A)\,T_A(t) - (1 - \epsilon_B)\,T_B(t) 
     + \epsilon^{\rm p}_A \, T^{\rm p}_A(t) - \epsilon^{\rm p}_B \, T^{\rm p}_B(t)
\label{eq:loss}
\eeq
where $\epsilon_A\ = \epsilon^{\rm p}_A + \epsilon^{\rm s}_A + \epsilon^{\rm
f}_A$ is the combined loss in the A-side optics: (p)rimary plus
(s)econdary mirrors, plus the (f)eed horn, and likewise for the B-side.
$T_{A,B}$ is the sky temperature in the direction of the A or B-side
line-of-sight; and $T^{\rm p}_{A,B}$ is the physical temperature of the A or
B-side primary mirror.

The first two terms are the sky signal attenuated by the overall loss in the A
and B side optics, respectively. The effects of loss imbalance, which arise when
$\epsilon_A \ne \epsilon_B$, have been studied extensively
\citep{jarosik/etal:2003, jarosik/etal:2007}.  We account for loss imbalance in
the data processing and we marginalize over residual uncertainties in the
imbalance coefficients when we form the pixel-pixel inverse covariance matrices
\citep{jarosik/etal:2007}. Updated estimates of the loss imbalance coefficients
based on fits to the 5 year data are reported in Table~\ref{tab:x_im}.

In the remainder of this section we focus on the last two emissive terms in
Equation~\ref{eq:loss}.  Recall that a \map\ differencing assembly consists of
two radiometers, 1 and 2, that are sensitive to orthogonal linear polarization
modes.  The temperature and polarization signals are extracted by forming the
sum and difference of the two radiometer outputs; thus, the emission terms we
need to evaluate are
\beq
d^{\rm p}_1(t) \pm d^{\rm p}_2(t)
  = \left(\frac{\epsilon^{\rm p}_{A1} \pm \epsilon^{\rm p}_{A2}}{1 - \epsilon}\right) \, T^{\rm p}_A
  - \left(\frac{\epsilon^{\rm p}_{B1} \pm \epsilon^{\rm p}_{B2}}{1 - \epsilon}\right) \, T^{\rm p}_B
\eeq
where $\epsilon^{\rm p}_{A1}$ is the A-side primary mirror emissivity measured
by radiometer 1, and so forth.  The factor of $1-\epsilon$ in the denominator
applies a small correction for the mean loss, $\epsilon \equiv (\epsilon_A +
\epsilon_B)/2$, and arises from the process of calibrating the data to a known
sky brightness temperature (\S\ref{sec:cal_improve}).  Note that we only pick up
a polarized response if $\epsilon_1 \ne \epsilon_2$.

\begin{figure}
\epsscale{1.0}
\plotone{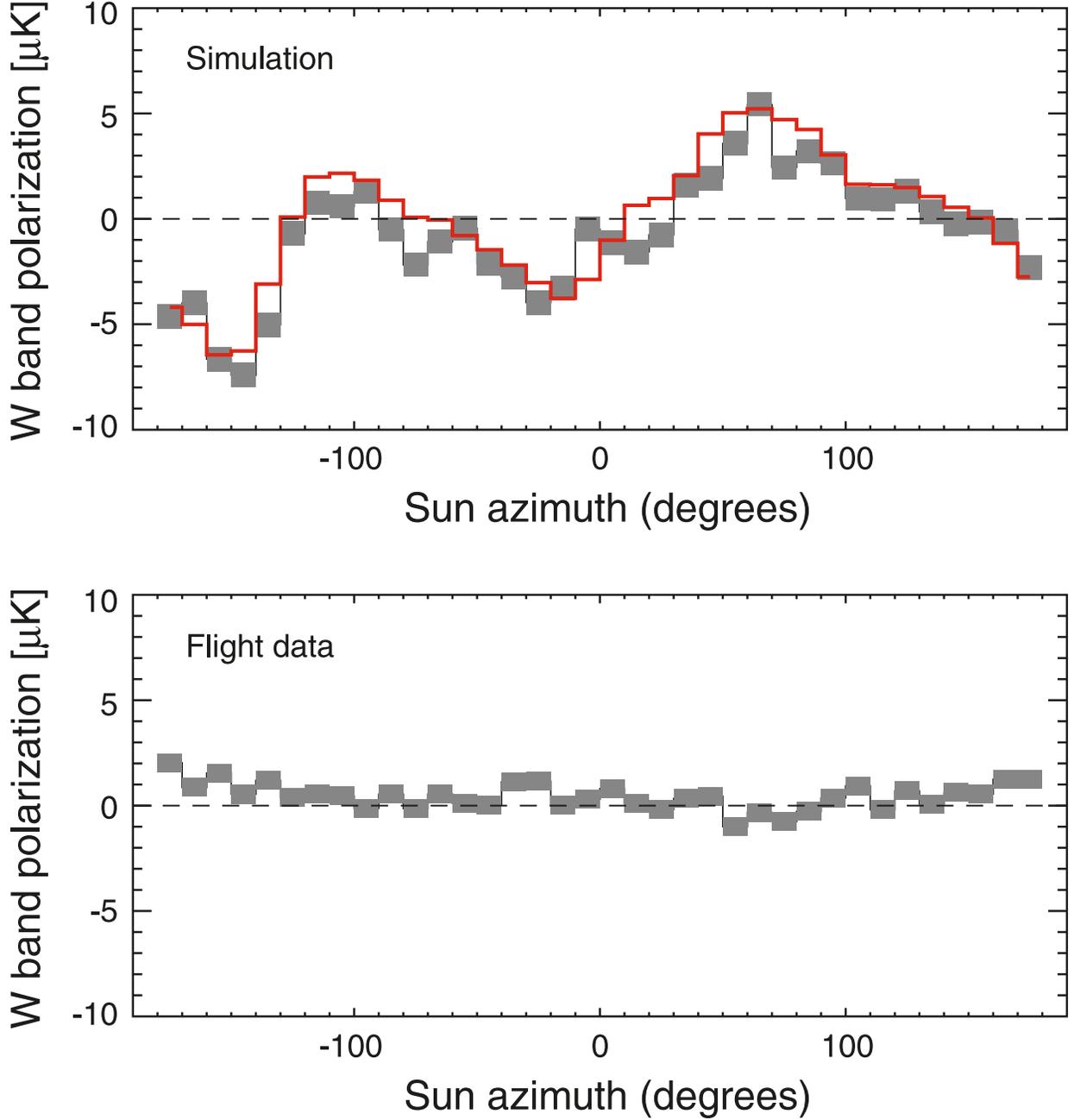}
\caption{{\it top}: Simulated W band data with a large polarized thermal
emission signal injected, binned by solar azimuth angle.  The red trace shows
the input waveform based on the flight mirror temperature profile and a model of
the polarized emissivity.  The black profile is the binned co-added data which
follows the input signal very well. The thickness of the points represents the 1
$\sigma$ uncertainty due to white noise.  {\it bottom}: Same as the top panel
but for the 5 year flight data.  The reduced $\chi^2$ of the binned data with
respect to zero is 2.1 for 36 degrees of freedom, but this does not account for
1/f noise, so the significance of this result requires further investigation. 
However, the much larger signal in the simulation did not produce an EE spectrum
with features present in the flight W band EE spectrum, so the feature in the
binned flight data cannot account for the excess $l=7$ emission.}
\label{fig:spin_bin}
\end{figure}

We have simulated this signal in the time-ordered data using the measured
primary mirror temperatures as template inputs.  The emissivity coefficients
were initially chosen to be consistent with the loss imbalance constraints. 
However, in order to produce a measurable polarization signal, we had to boost
the emissivity differences to the point where they became unphysical, that is $|
\epsilon_1 - \epsilon_2 | > | \epsilon_1 + \epsilon_2 |$.  Nonetheless, it was
instructive to analyze this simulation by binning the resulting data (which also
includes sky signal and noise) as a function of solar azimuth.  The results are
shown in the top panel of Figure~\ref{fig:spin_bin} which shows 3 years of
co-added W band polarization data, the $d_1-d_2$ channel; the input emissive
signal is shown in red for comparison.  We are clearly able to detect such a
signal with this manner of binning.  We also computed the low-$l$ polarization
spectra and found that, despite the large spin modulated input signal, the
signal induced in the power spectrum  was less than 2 $\mu$K$^2$ in
$l(l+1)C^{EE}_l/2\pi$, which is insufficient to explain the $l=7$ feature in the
W band EE spectrum.

In parallel with the simulation analysis, we have binned the flight radiometer
data by solar azimuth angle to search for spin modulated features in the
polarization data.  The results for W band are shown in the bottom panel of
Figure~\ref{fig:spin_bin} for the 5 year data.  While the $\chi^2$ per degree of
freedom relative to zero is slightly high, there is no compelling evidence for a
coherent spin modulated signal at the $\sim$2 $\mu$K level.  In contrast, the
simulation yielded spin modulated signals of 5-10 $\mu$K and still failed to
produce a significant effect in the EE spectrum.  Hence we conclude that thermal
emission from the \map\ optics cannot explain the excess W band EE signal.  In
any event, we continue to monitor the spin modulated data for the emergence of a
coherent signal.

\subsection{Ka Band Tests}
\label{sec:pol_Ka}

The analysis presented in \S\ref{sec:pol_year} shows that the Ka band
polarization data is comparable to the Q and V band data in its internal
consistency.  That analysis was performed on data that had been foreground
cleaned using the template method discussed in \citet{page/etal:2007} and
updated in \citet{gold/etal:prep}.  In order to assess whether or not this
cleaned Ka band data is suitable for use in cosmological parameter estimation we
subject it to two further tests: 1) a null test in which Ka band data is
compared to the combined Q and V band data, and 2) a parameter estimation based
solely on Ka band data.

\begin{figure}
\epsscale{1.0}
\plotone{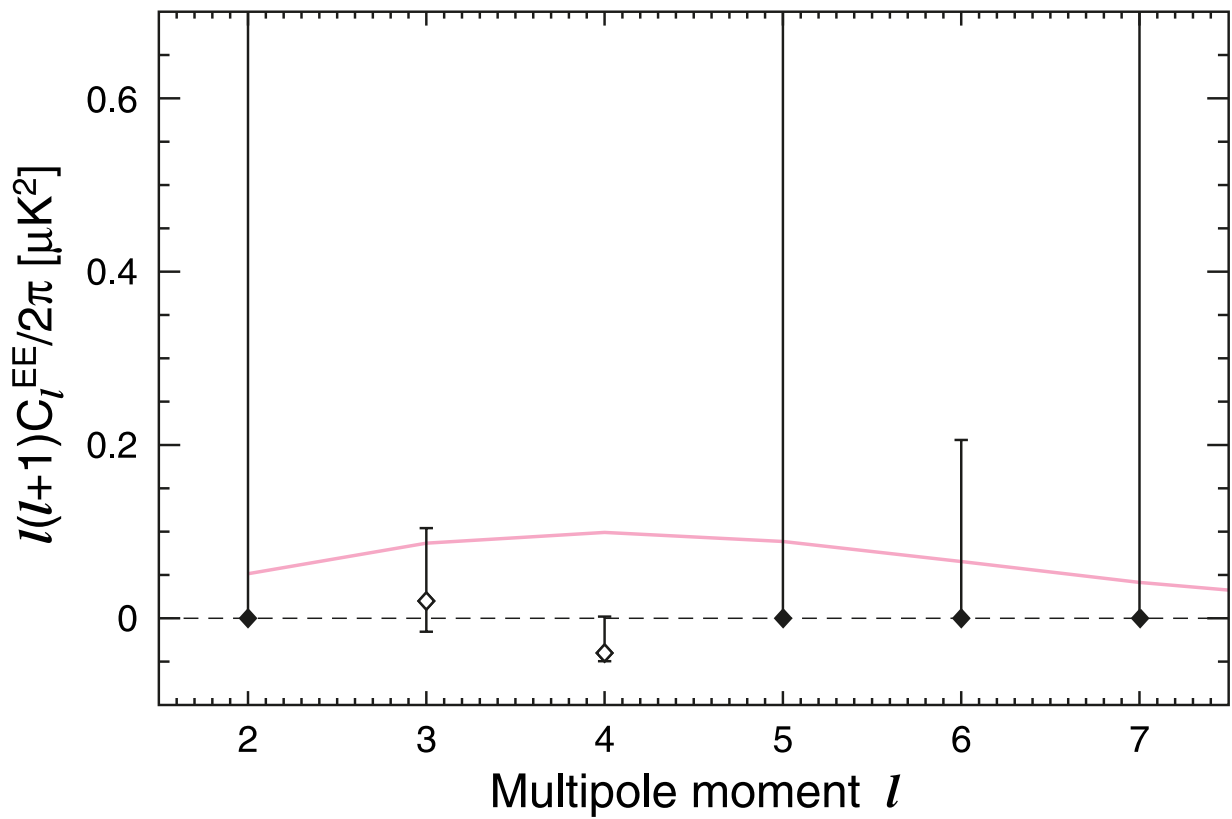}
\caption{The EE power spectrum computed from the null sky maps,
$\frac{1}{2}S_{\rm Ka} - \frac{1}{2}S_{\rm QV}$, where S = Q,U are the
polarization Stokes parameters, and $S_{\rm QV}$ is the optimal combination of
the Q and V band data.  The pink curve is the best-fit theoretical spectrum
from \citet{dunkley/etal:prep}.  The spectrum derived from the null maps is
consistent with zero.}
\label{fig:null_EE}
\end{figure}

\begin{figure}
\epsscale{1.0}
\plotone{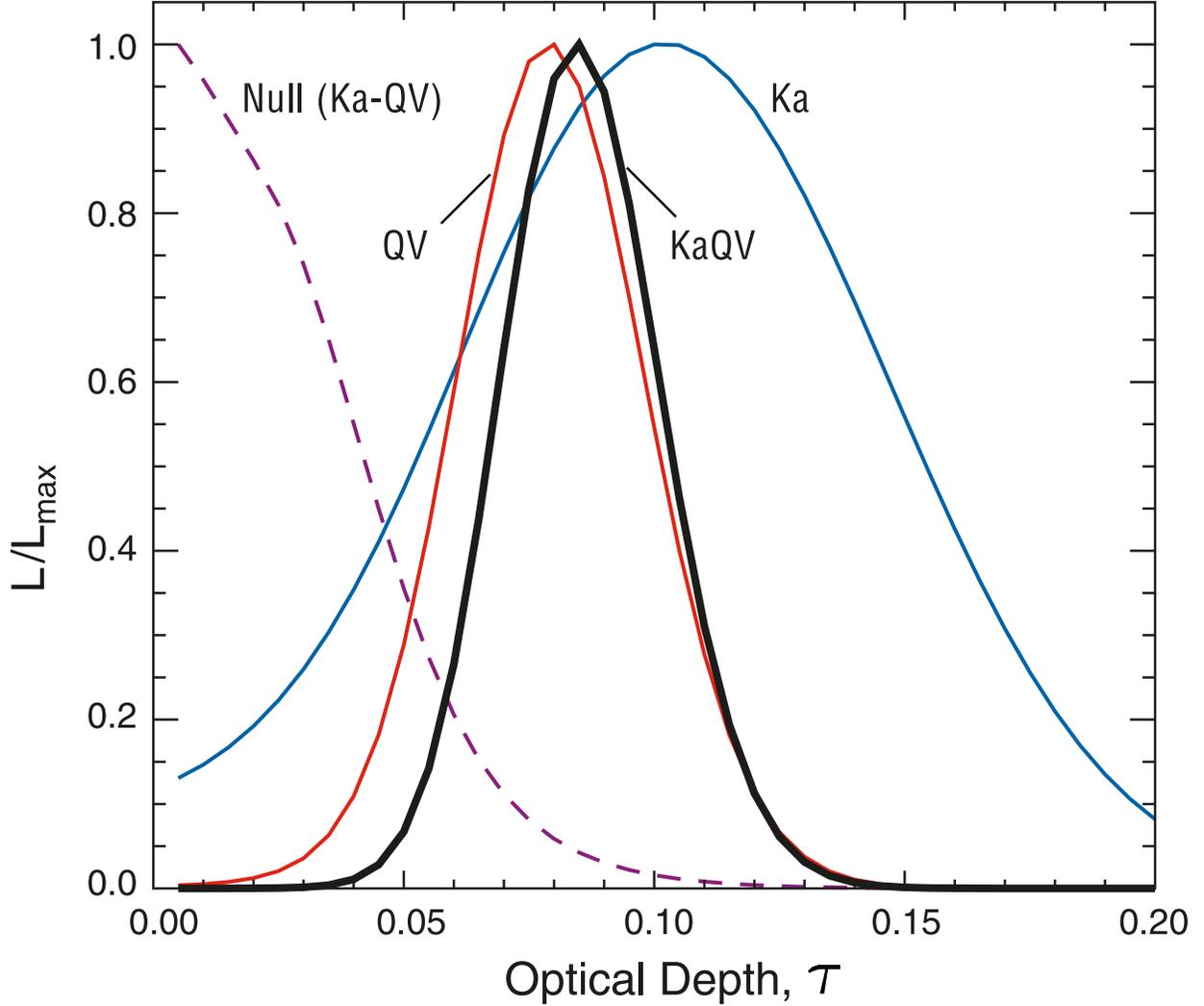}
\caption{Estimates of the optical depth from a variety of data combinations. 
The dashed curve labeled Null uses the same null sky maps used in
Figure~\ref{fig:null_EE}.  The optical depth obtained from Ka band data alone
(blue) is consistent with independent estimates from the combined Q and V band
data (red).  The final 5 year analysis uses Ka, Q, and V band data combined
(black). These estimates all use a 1-parameter likelihood estimation, holding
other parameters fixed except for the fluctuation amplitude, which is adjusted
to fit the first acoustic peak in the TT spectrum \citep{page/etal:2007}. The
degeneracy between $\tau$ and other $\Lambda$CDM parameters is small: see
Figure~7 of \citet{dunkley/etal:prep}.}
\label{fig:tau_test}
\end{figure}

For the null test, we form polarization maps by taking differences, 
$\frac{1}{2}S_{\rm Ka} - \frac{1}{2}S_{\rm QV}$, where S = Q,U are the
polarization Stokes parameters, $S_{\rm Ka}$ are the maps formed from the Ka
band data, and $S_{\rm QV}$ are the maps formed from the optimal combination of
the Q and V band data.   We evaluate the EE power spectrum from these null maps
by evaluating the likelihood mode by mode while holding the other multipoles
fixed at zero.  The results are shown in Figure~\ref{fig:null_EE}, along with
the best-fit model spectrum based on the final 5 year $\Lambda$CDM analysis. 
The spectrum is clearly consistent with zero, but to get a better sense of the
power of this test, we have also used these null maps to estimate the optical
depth parameter, $\tau$.  The result of that analysis is shown as the dashed
curve in Figure~\ref{fig:tau_test}, where we find that the null likelihood peaks
at $\tau=0$ and excludes the most-likely cosmological value with $\sim$95\%
confidence.

As a separate test, we evaluate the $\tau$ likelihood using only the
template-cleaned Ka band signal maps.  The result of that test is shown as the
blue curve in Figure~\ref{fig:tau_test}.  While the uncertainty in the Ka band
estimate is considerably larger than the combined QV estimate (shown in red),
the estimates are highly consistent.  The result of combining Ka, Q, and V band
data is shown in the black curve.  

\citet{dunkley/etal:prep} present a complementary method of foreground cleaning
that makes use of Ka band data, in conjunction with K, Q, and V band data. 
Using a full 6 parameter likelihood evalutaion, they compare the optical depth
inferred from the two cleaning methods while using the full combined data sets
in both cases: see Figure~9 of \citet{dunkley/etal:prep} for details.  Based
on these tests, we conclude that the Ka band data is sufficiently free of
systematic errors and residual foreground signals that it is suitable for
cosmological studies.  The use of this band significantly enhances the overall
polarization sensitivity of \map.

\section{SUMMARY OF 5-YEAR SCIENCE RESULTS}
\label{sec:science}

\begin{figure}
\epsscale{0.80}
\plotone{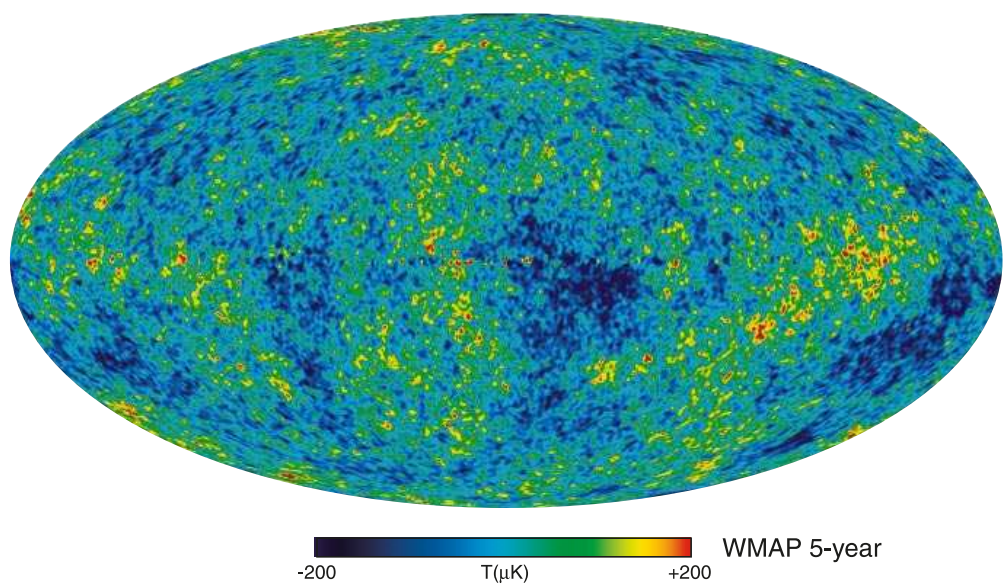}
\caption{The foreground-reduced Internal Linear Combination (ILC) map based on
the 5 year \map\ data.}
\label{fig:ilc_map}
\end{figure}

Detailed presentations of the scientific results from the 5 year data are given
by \citet{gold/etal:prep}, \citet{wright/etal:prep}, \citet{nolta/etal:prep},
\citet{dunkley/etal:prep}, and \citet{komatsu/etal:prep}.  Starting with the 5
year temperature and polarization maps, with their improved calibration,
\citet{gold/etal:prep} give a new Markov Chain Monte Carlo-based analysis of
foreground emission in the data.  Their results are broadly consistent with
previous analyses by the \map\ team and others \citep{eriksen/etal:2007}, while
providing some new results on the microwave spectra of bright sources in the
Galactic plane that aren't well fit by simple power-law foreground models. 
Figure~\ref{fig:ilc_map} shows the 5 year CMB map based on the internal linear
combination (ILC) method of foreground removal.  

\citet{wright/etal:prep} give a comprehensive analysis of the extragalactic
sources in the 5 year data, including a new analysis of variability made
possible by the multi-year coverage.  The 5 year \map\ source catalog now
contains 390 objects and is reasonably complete to a flux of 1 Jy away from the
Galactic plane.  The new analysis of the \map\ beam response
\citep{hill/etal:prep} has led to more precise estimates of the point source
flux scale for all 5 \map\ frequency bands.  This information is incorporated in
the new source catalog \citep{wright/etal:prep}, and is also used to provide new
brightness estimates of Mars, Jupiter, and Saturn \citep{hill/etal:prep}.  We
find significant (and expected) variability in Mars and Saturn over the course
of 5 years and use that information to provide a preliminary recalibration of a
Mars brightness model \citep{wright:2007}, and to fit a simple model of Saturn's
brightness as a function of ring inclination.

\begin{figure}
\epsscale{1.0}
\plotone{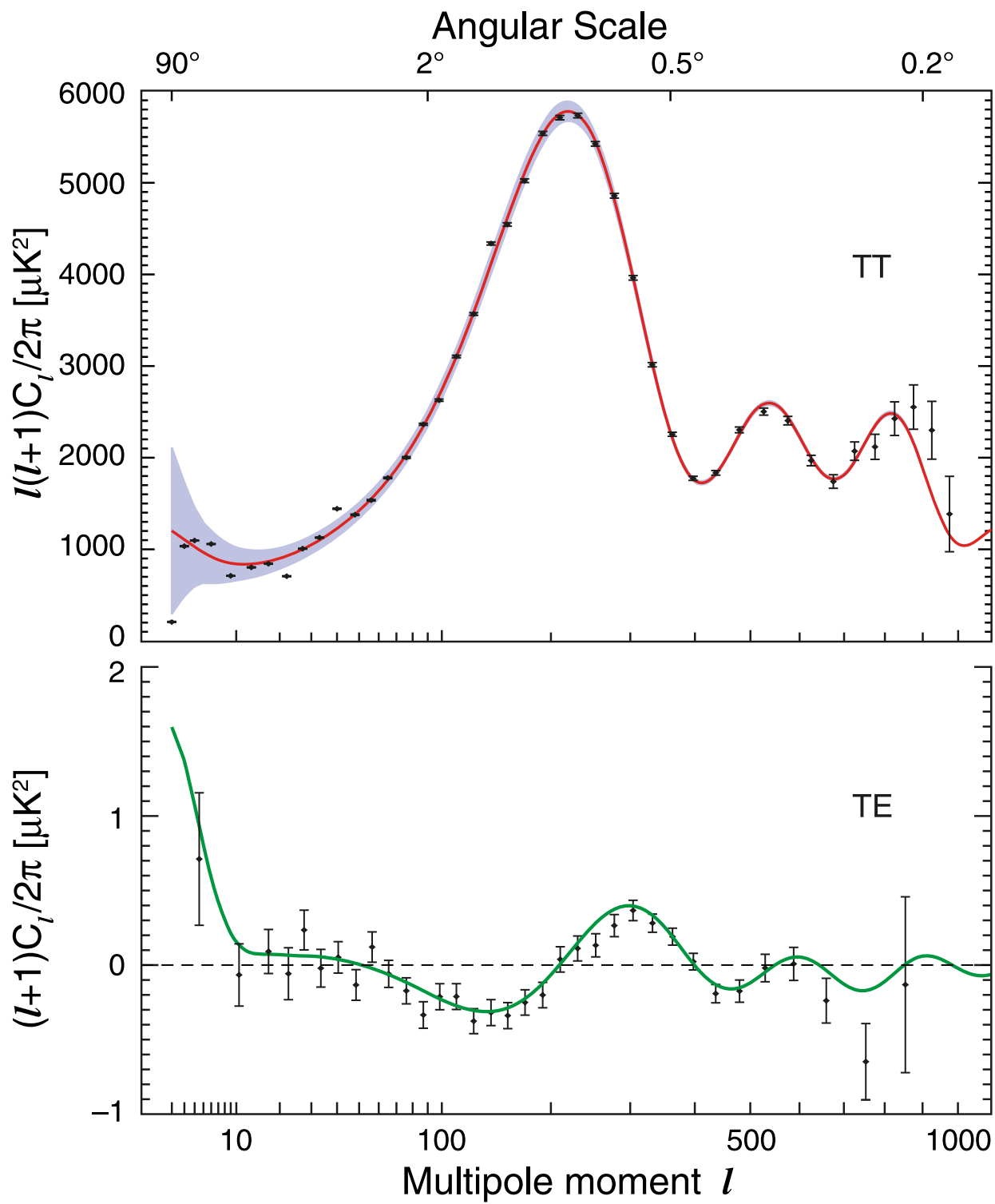}
\caption{The temperature (TT) and temperature-polarization correlation (TE)
power spectra based on the 5 year \map\ data.  The addition of 2 years of data
provide more sensitive measurements of the third peak in TT and the high-$l$ TE
spectrum, especially the second trough.}
\label{fig:tt_te}
\end{figure}

The temperature and polarization power spectra are presented in
\citet{nolta/etal:prep}.  The spectra are all consistent with the 3 year results
with improvements in sensitivity commensurate with the additional integration
time.  Further improvements in our understanding of the absolute calibration and
beam response have allowed us to place tighter uncertainties on the power
spectra, over and above the reductions from additional data.  These changes are
all reflected in the new version of the \map\ likelihood code.  The most notable
improvements arise in the third acoustic peak of the TT spectrum, and in all of
the polarization spectra; for example, we now see unambiguous evidence for a 2nd
dip in the high-$l$ TE spectrum, which further constrains deviations from the
standard $\Lambda$CDM model.  The 5 year TT and TE spectra are shown in
Figure~\ref{fig:tt_te}.  We have also generated new maximum likelihood estimates
of the low-$l$ polarization spectra: TE, TB, EE, EB, and BB to complement our
earlier estimates based on pseudo-$C_l$ methods \citep{nolta/etal:prep}.  The
TB, EB, and BB spectra remain consistent with zero.

The cosmological implications of the 5 year \map\ data are discussed in detail
in \citet{dunkley/etal:prep} and \citet{komatsu/etal:prep}.  The now-standard
cosmological model: a flat universe dominated by vacuum energy and dark matter,
seeded by nearly scale-invariant, adiabatic, Gaussian random-phase fluctuations,
continues to fit the 5 year data.  \map\ has now determined the key parameters
of this model to high precision; a summary of the 5 year parameter results is
given in Table~\ref{tab:best_param}.  The most notable improvements are the
measurements of the dark matter density, $\Omega_c h^2$, and the amplitude of
matter fluctuations today, $\sigma_8$.  The former is determined with 6\%
uncertainty using \map\ data only \citep{dunkley/etal:prep}, and with 3\%
uncertainty when \map\ data is combined with BAO and SNe constraints
\citep{komatsu/etal:prep}.  The latter is measured to 5\% with \map\ data, and
to 3\% when combined with other data.  The redshift of reionization is
\ensuremath{z_{\rm reion}} = \ensuremath{11.0\pm 1.4}, if
the universe were reionized instantaneously.  The 2 $\sigma$ lower limit is
$z_{\rm reion} \gt 8.2$, and instantaneous reionization at $z_{\rm reion} = 6$
is rejected at 3.5 $\sigma$.  The \map\ data continues to favor models with a
tilted primordial spectrum, 
\ensuremath{n_s} = \ensuremath{0.963^{+ 0.014}_{- 0.015}}.  
\citet{dunkley/etal:prep} discuss how the $\Lambda$CDM  model continues to fit a
host of other astronomical data as well.

Moving beyond the standard $\Lambda$CDM model, when \map\ data is combined with
BAO and SNe observations \citep{komatsu/etal:prep}, we find no evidence for
running in the spectral index of scalar fluctuations,
\ensuremath{dn_s/d\ln{k} = -0.028\pm 0.020} (68\% CL).  
The new limit on the tensor-to-scalar ratio is
\ensuremath{r < 0.22\ \mbox{(95\% CL)}}, 
and we obtain tight, simultaneous limits on the (constant) dark energy equation
of state and the spatial curvature of the universe: 
\ensuremath{-0.14<1+w<0.12\ \mbox{(95\% CL)}}
and
\ensuremath{-0.0179<\Omega_k<0.0081\ \mbox{(95\% CL)}}. 
The angular power spectrum now exhibits the signature of the cosmic neutrino
background: the number of relativistic degrees of freedom, expressed in units of
the effective number of neutrino species, is found to be
\ensuremath{N_{\rm eff} = 4.4\pm 1.5} (68\% CL), 
consistent with the standard value of 3.04.  Models with $N_{\rm eff} = 0$ are
disfavored at $\gt$99.5\% confidence.  A summary of the key cosmological
parameter values is given in Table~\ref{tab:best_param}, where we provide
estimates using \map\ data alone and \map\ data combined with BAO and SNe
observations.  A complete tabulation of all parameter values for each model and
dataset combination we studied is available on LAMBDA.

The new data also place more stringent limits on deviations from Gaussianity,
parity violations, and the amplitude of isocurvature fluctuations
\citep{komatsu/etal:prep}. For example, new limits on physically motivated
primordial non-Gaussianity parameters are $-9 < \fnlKS <111$ (95\% CL) and $-151
< \fnleq < 253$ (95\% CL) for the local and equilateral models, respectively.

\section{CONCLUSIONS}
\label{sec:conclusions}

We have presented an overview of the 5 year \map\ data and have highlighted the
improvements we have made to the data processing and analysis since the 3 year
results were presented.  The most substantive improvements to the processing
include a new method for establishing the absolute gain calibration (with
reduced uncertainty), and a more complete analysis of the \map\ beam response
made possible by additional data and a higher fidelity physical optics model. 
Numerous other processing changes are outlined in \S\ref{sec:change}.

The 5 year sky maps are consistent with the 3 year maps and have noise levels
that are $\sqrt{5}$ times less than the single year maps.  The new maps are
compared to the 3 year maps in \S\ref{sec:maps}.  The main changes to the
angular power spectrum are as follows: at low multipoles ($l \la 100$) the
spectrum is $\sim$0.5\% higher than the 3 year spectrum (in power units) due to
the new absolute gain determination.  At higher multipoles it is increased by
$\sim$2.5\%, due to the new beam response profiles, as explained in
\S\ref{sec:beams} and in \citet{hill/etal:prep}.  These changes are consistent
with the 3 year uncertainties when one accounts for both the 0.5\% gain
uncertainty (in temperature units) and the 3 year beam uncertainties, which were
incorporated into the likelihood code.

We have applied a number of new tests to the polarization data to check internal
consistency and to look for new systematic effects in the W band data
(\S\ref{sec:pol_test}).  As a result of these tests, and of new analyses of
polarized foreground emission \citep{dunkley/etal:prep}, we have concluded that
Ka band data can be used along with Q and V band data for cosmological
analyses.  However, we still find a number of features in the W band
polarization data that preclude its use, except in the Galactic plane where the
signal to noise is relatively high.  We continue to investigate the causes of
this and have identified new clues to follow up on in future studies
(\S\ref{sec:pol_year}).

Scientific results gathered from the suite of 5 year papers are summarized in
\S\ref{sec:science}.  The highlights include smaller uncertainties in the
optical depth, $\tau$, due to a combination of additional years of data and to
the inclusion of Ka band polarization data: instantaneous reionization at
$z_{\rm reion} = 6$ is now rejected at 3.5 $\sigma$.  New evidence favoring a
non-zero neutrino abundance at the epoch of last scattering, made possible by
improved measurements of the third acoustic peak; and new limits on the
nongaussian parameter $f_{NL}$, based on additional data and the application of
a new, more optimal bispectrum estimator.  The 5 year data continue to favor a
tilted primordial fluctuation spectrum, in the range $n_s \sim 0.96$, but a
purely scale invariant spectrum cannot be ruled out at $\gt$3 $\sigma$
confidence.  

The \map\ observatory continues to operate at L2 as designed, and the addition
of two years of flight data has allowed us to make significant advances in
characterizing the instrument.  Additional data beyond 5 years will give us a
better understanding of the instrument, especially with regards to the W band
polarization data since the number of jackknife combinations scales like the
square of the number of years of operation.  If W band data can be incorporated
into the EE power spectrum estimate, it would become possible to constrain a
second reionization parameter and thereby further probe this important epoch in
cosmology.  The \map\ data continues to uphold the standard $\Lambda$CDM model
but more data may reveal new surprises.

\section{DATA PRODUCTS}
\label{sec:data}

All of the \map\ data is released to the research community for further analysis
through the Legacy Archive for Microwave Background Data Analysis (LAMBDA) at
http://lambda.gsfc.nasa.gov.  The products include the complete 5 year
time-ordered data archive (both raw and calibrated); the calibrated sky maps in
a variety of processing stages (single year by DA, multi-year by band, high
resolution and low resolution, smoothed, foreground-subtracted, and so forth);
the angular power spectra and cosmological model likelihood code; a full table
of model parameter values for a variety of model and data sets (including the
best-fit model spectra and Markov chains); and a host of ancillary data to
support further analysis.  The \map\ Explanatory Supplement provides detailed
information about the \map\ in-flight operations and data products
\citep{limon/etal:prep}.

\acknowledgements

The \map\ mission is made possible by the support of the Science Mission
Directorate Office at NASA Headquarters.  This research was additionally
supported by NASA grants NNG05GE76G, NNX07AL75G S01, LTSA03-000-0090,
ATPNNG04GK55G, and ADP03-0000-092.  EK acknowledges support from an Alfred P.
Sloan Research Fellowship.  This research has made use of NASA's Astrophysics
Data System Bibliographic Services.  We acknowledge use of the HEALPix, CAMB,
CMBFAST, and CosmoMC packages.

\bibliographystyle{wmap}
\bibliography{wmap}

\begin{thebibliography}{49}
\expandafter\ifx\csname natexlab\endcsname\relax\def\natexlab#1{#1}\fi

\bibitem[{{Astier} et~al.(2006)}]{astier/etal:2006}
{Astier}, P., et~al. 2006, \aap, 447, 31

\bibitem[{{Barnes} et~al.(2003)}]{barnes/etal:2003}
{Barnes}, C., et~al. 2003, \apjs, 148, 51

\bibitem[{{Bennett} et~al.(2003)}]{bennett/etal:2003}
{Bennett}, C.~L., et~al. 2003, \apj, 583, 1

\bibitem[{{Beno{\^i}t} et~al.(2003)}]{benoit/etal:2003b}
{Beno{\^i}t}, A., et~al. 2003, \aap, 399, L25

\bibitem[{Born \& Wolf(1980)}]{born/wolf:POO:6e}
Born, M. \& Wolf, E. 1980, Principles of Optics, sixth edn. (Pergamon Press)

\bibitem[{{Cole} et~al.(2005)}]{cole/etal:2005}
{Cole}, S., et~al. 2005, \mnras, 362, 505

\bibitem[{{Copi} et~al.(2007){Copi}, {Huterer}, {Schwarz}, \&
  {Starkman}}]{copi/etal:2007}
{Copi}, C.~J., {Huterer}, D., {Schwarz}, D.~J., \& {Starkman}, G.~D. 2007,
  \prd, 75, 023507

\bibitem[{Creminelli et~al.(2006)Creminelli, Nicolis, Senatore, Tegmark, \&
  Zaldarriaga}]{creminelli/etal:2006}
Creminelli, P., Nicolis, A., Senatore, L., Tegmark, M., \& Zaldarriaga, M.
  2006, JCAP, 0605, 004

\bibitem[{{Dunkley} et~al.(2008)}]{dunkley/etal:prep}
{Dunkley}, J., et~al. 2008, ArXiv e-prints, 803

\bibitem[{{Eisenstein} et~al.(2005)}]{eisenstein/etal:2005}
{Eisenstein}, D.~J., et~al. 2005, \apj, 633, 560

\bibitem[{{Eriksen} et~al.(2007{\natexlab{a}}){Eriksen}, {Banday},
  {G{\'o}rski}, {Hansen}, \& {Lilje}}]{eriksen/etal:2007c}
{Eriksen}, H.~K., {Banday}, A.~J., {G{\'o}rski}, K.~M., {Hansen}, F.~K., \&
  {Lilje}, P.~B. 2007{\natexlab{a}}, \apjl, 660, L81

\bibitem[{{Eriksen} et~al.(2007{\natexlab{b}}){Eriksen}, {Jewell}, {Dickinson},
  {Banday}, {Gorski}, \& {Lawrence}}]{eriksen/etal:prep}
{Eriksen}, H.~K., {Jewell}, J.~B., {Dickinson}, C., {Banday}, A.~J., {Gorski},
  K.~M., \& {Lawrence}, C.~R. 2007{\natexlab{b}}, ArXiv e-prints, 709

\bibitem[{{Eriksen} et~al.(2007{\natexlab{c}})}]{eriksen/etal:2007}
{Eriksen}, H.~K., et~al. 2007{\natexlab{c}}, \apj, 656, 641

\bibitem[{{Freedman} et~al.(2001)}]{freedman/etal:2001}
{Freedman}, W.~L., et~al. 2001, \apj, 553, 47

\bibitem[{{Gold} et~al.(2008)}]{gold/etal:prep}
{Gold}, B. et~al. 2008, \apjs

\bibitem[{Gorski et~al.(2005)Gorski, Hivon, Banday, Wandelt, Hansen, Reinecke,
  \& Bartlemann}]{gorski/etal:2005}
Gorski, K.~M., Hivon, E., Banday, A.~J., Wandelt, B.~D., Hansen, F.~K.,
  Reinecke, M., \& Bartlemann, M. 2005, \apj, 622, 759

\bibitem[{{Halverson} et~al.(2002)}]{halverson/etal:2002}
{Halverson}, N.~W., et~al. 2002, \apj, 568, 38

\bibitem[{{Hill} et~al.(2008)}]{hill/etal:prep}
{Hill}, R. et~al. 2008, \apjs

\bibitem[{{Hinshaw} et~al.(2003)}]{hinshaw/etal:2003b}
{Hinshaw}, G., et~al. 2003, \apjs, 148, 63

\bibitem[{{Hinshaw} et~al.(2007)}]{hinshaw/etal:2007}
---. 2007, \apjs, 170, 288

\bibitem[{{Hunt} \& {Sarkar}(2007)}]{hunt/sarkar:2007}
{Hunt}, P. \& {Sarkar}, S. 2007, \prd, 76, 123504

\bibitem[{{Jarosik} et~al.(2003)}]{jarosik/etal:2003}
{Jarosik}, N., et~al. 2003, \apjs, 145, 413

\bibitem[{{Jarosik} et~al.(2007)}]{jarosik/etal:2007}
---. 2007, \apjs, 170, 263

\bibitem[{{Kogut} et~al.(1996)}]{Kogut/etal:1996d}
{Kogut}, A., et~al. 1996, \apj, 470, 653

\bibitem[{{Komatsu} et~al.(2008)}]{komatsu/etal:prep}
{Komatsu}, E., et~al. 2008, ArXiv e-prints, 803

\bibitem[{{Land} \& {Magueijo}(2007)}]{land/magueijo:2007}
{Land}, K. \& {Magueijo}, J. 2007, \mnras, 378, 153

\bibitem[{Lee et~al.(2001)}]{lee/etal:2001}
Lee, A.~T. et~al. 2001, \apjl, 561, L1

\bibitem[{{Limon} et~al.(2008)}]{limon/etal:prep}
{Limon}, M., et~al. 2008, Wilkinson Microwave Anisotropy Probe ({\sl WMAP}):
  Explanatory Supplement,
  \texttt{http://lambda.gsfc.nasa.gov/data/map/doc/MAP\_supplement.pdf}

\bibitem[{{Mather} et~al.(1999){Mather}, {Fixsen}, {Shafer}, {Mosier}, \&
  {Wilkinson}}]{mather/etal:1999}
{Mather}, J.~C., {Fixsen}, D.~J., {Shafer}, R.~A., {Mosier}, C., \&
  {Wilkinson}, D.~T. 1999, \apj, 512, 511

\bibitem[{Miller et~al.(1999)}]{miller/etal:1999}
Miller, A.~D. et~al. 1999, \apjl, 524, L1

\bibitem[{Netterfield et~al.(2002)}]{netterfield/etal:2002}
Netterfield, C.~B. et~al. 2002, \apj, 571, 604

\bibitem[{{Nolta} et~al.(2008)}]{nolta/etal:prep}
{Nolta}, M.~R. et~al. 2008, \apjs

\bibitem[{{Page} et~al.(2007)}]{page/etal:2007}
{Page}, L., et~al. 2007, \apjs, 170, 335

\bibitem[{{Pearson} et~al.(2003)}]{pearson/etal:2003}
{Pearson}, T.~J., et~al. 2003, \apj, 591, 556

\bibitem[{{Peiris} et~al.(2003)}]{peiris/etal:2003}
{Peiris}, H.~V., et~al. 2003, \apjs, 148, 213

\bibitem[{{Percival} et~al.(2007){Percival}, {Cole}, {Eisenstein}, {Nichol},
  {Peacock}, {Pope}, \& {Szalay}}]{percival/etal:2007c}
{Percival}, W.~J., {Cole}, S., {Eisenstein}, D.~J., {Nichol}, R.~C., {Peacock},
  J.~A., {Pope}, A.~C., \& {Szalay}, A.~S. 2007, \mnras, 381, 1053

\bibitem[{{Percival} et~al.(2001)}]{percival/etal:2001}
{Percival}, W.~J., et~al. 2001, \mnras, 327, 1297

\bibitem[{{Riess} et~al.(2007)}]{riess/etal:2007}
{Riess}, A.~G., et~al. 2007, \apj, 659, 98

\bibitem[{{Scott} et~al.(2003)}]{scott/etal:2003}
{Scott}, P.~F., et~al. 2003, \mnras, 341, 1076

\bibitem[{{Shafieloo} \& {Souradeep}(2007)}]{shafieloo/souradeep:2007}
{Shafieloo}, A. \& {Souradeep}, T. 2007, ArXiv e-prints, 709

\bibitem[{{Spergel} et~al.(2003)}]{spergel/etal:2003}
{Spergel}, D.~N., et~al. 2003, \apjs, 148, 175

\bibitem[{{Spergel} et~al.(2007)}]{spergel/etal:2007}
---. 2007, \apjs, 170, 377

\bibitem[{{Tegmark} et~al.(2004)}]{tegmark/etal:2004a}
{Tegmark}, M., et~al. 2004, \prd, 69, 103501

\bibitem[{{Tegmark} et~al.(2006)}]{tegmark/etal:2006}
---. 2006, \prd, 74, 123507

\bibitem[{{Wood-Vasey} et~al.(2007)}]{wood-vasey/etal:2007}
{Wood-Vasey}, W.~M., et~al. 2007, \apj, 666, 694

\bibitem[{{Wright}(2007)}]{wright:2007}
{Wright}, E.~L. 2007, ArXiv Astrophysics e-prints

\bibitem[{{Wright} et~al.(2008)}]{wright/etal:prep}
{Wright}, E.~L. et~al. 2008, \apjs

\bibitem[{{Yadav} et~al.(2007){Yadav}, {Komatsu}, {Wandelt}, {Liguori},
  {Hansen}, \& {Matarrese}}]{yadav/etal:prep}
{Yadav}, A.~P.~S., {Komatsu}, E., {Wandelt}, B.~D., {Liguori}, M., {Hansen},
  F.~K., \& {Matarrese}, S. 2007, ArXiv e-prints, 711

\bibitem[{{Yadav} \& {Wandelt}(2008)}]{yadav/wandelt:2008}
{Yadav}, A.~P.~S. \& {Wandelt}, B.~D. 2008, Physical Review Letters, 100,
  181301

\end{thebibliography}

\appendix

\section{FISHER MATRIX ANALYSIS OF CALIBRATION AND SKY MAP FITS}
\label{app:cal_fisher}

\subsection{Least Squares Calibration and Sky Model Fitting}

Let $i$ be a time index in the time ordered data.  Let $g^j$ be
parameters for the gain, $a_{l m}$ be parameters for the
temperature anisotropy and $b^k$ be parameters for the baseline
offset.

The model of the time-ordered data (TOD) is
\beq
m_i = g_i \left[\Delta T_{vi} + \Delta T_{ai} \right] + b_i,
\eeq
where $i$ is a time index, $\Delta T_{vi}$ is the differential dipole signal at
time step $i$, including the CMB dipole, and $\Delta T_{ai}$ is the differential
anisotropy signal at time step $i$.  The parameters of the model are the hourly
gain and baseline values, and the sky map pixel temperatures (which goes into
forming $\Delta T_a$.  We fit for them by minimizing
\beq
\chi^2 = \sum_i \frac{(c_i - m_i)^2}{\sigma_i^2},
\eeq
where $c_i$ is the raw data, in counts, and $\sigma_i$ is the {\em rms} of the
$i$th observation, in counts. The Fisher matrix requires taking the second
derivative of $\chi^2$ with respect to all parameters being fit.  In order to
reduce the dimensionality of the problem to something manageable, we expand the
calibration and sky signal in terms of a small number of parameters.  We can
write
\beqa
g_i & = & \sum_j g^j G_{ji}, \\
b_i & = & \sum_k b^k B_{ki}, \\
\Delta T_{ai} & = & \sum_{lm} a_{lm} \left[Y_{lm}(\hat{n}_{Ai}) 
                                    - Y_{lm}(\hat{n}_{Bi})\right],
\eeqa
where $G$ and $B$ are function of time (defined below), $a_{lm}$ are the
harmonic coefficients of the map, and $\hat{n}_{Ai}$ is the unit vector of the
$A$-side feed at time step $i$, and likewise for $B$.

A reasonable set of basis functions for the gain and baseline allow for an
annual modulation and a small number of higher harmonics.  Note that this does
not include power at the spin or precession period, which might be an important
extension to consider.  For now we consider the trial set
\beq
G_{ji} = \left\{
\begin{array}{ll}
1 & j=0 \\
\cos j\theta_i & j=1,\ldots,j_{\rm max} \\
\sin (j-j_{\rm max})\theta_i & j=j_{\rm max}+1,\ldots,2j_{\rm max}
\end{array}
\right. ,
\eeq
and
\beq
B_{ki} = \left\{
\begin{array}{ll}
1 & k=0 \\
\cos k\theta_i & k=1,\ldots,k_{\rm max} \\
\sin (k-k_{\rm max})\theta_i & k=k_{\rm max}+1,\ldots,2k_{\rm max}
\end{array}
\right. ,
\eeq
where $\theta = \tan^{-1}(\hat{n}_y/\hat{n}_x)$.  Here $\hat{n}$ is the unit
vector from \map\ to the Sun, and the components are evaluated in ecliptic
coordinates.

\subsection{Evaluation of the Fisher Matrix}

We wish to evaluate the 2nd derivative
\beq
\frac{1}{2}\,\frac{\partial^2 \chi^2}{\partial p_i \partial p_j}
\eeq
where $p_i$ and $p_j$ are the parameters we are trying to fit.  The needed first
derivatives are
\beq
\frac{1}{2}\,\frac{\partial \chi^2}{\partial g^{j'}} 
 = -\sum_i \frac{(c_i - m_i) G_{j'i} \left[\Delta T_{vi} + \Delta T_{ai}\right]}
   {\sigma_i^2},
\eeq
\beq
\frac{1}{2}\,\frac{\partial \chi^2}{\partial b^{k'}} 
 = -\sum_i \frac{(c_i - m_i) B_{k'i}}
   {\sigma_i^2},
\eeq
\beq
\frac{1}{2}\,\frac{\partial \chi^2}{\partial a_{l'm'}}
 = -\sum_i \frac{(c_i - m_i) g_i 
   \left[Y_{l'm'}(\hat{n}_{Ai}) - Y_{l'm'}(\hat{n}_{Bi})\right]}
   {\sigma_i^2}.
\eeq
Then
\beq
\frac{1}{2}\,\frac{\partial^2 \chi^2}{\partial g^{j'} \partial g^{j''}} 
 = \sum_i \frac{G_{j'i} \left[\Delta T_{vi} + \Delta T_{ai}\right]
              \,G_{j''i} \left[\Delta T_{vi} + \Delta T_{ai}\right]}
   {\sigma_i^2}
\eeq
\beq
\frac{1}{2}\,\frac{\partial^2 \chi^2}{\partial g^{j'} \partial a_{l'm'}} 
 = \sum_i \frac{g_i \left[Y_{l'm'}(\hat{n}_{Ai}) - Y_{l'm'}(\hat{n}_{Bi})\right]
          G_{j'i} \left[\Delta T_{v i} + \Delta T_{ai}\right]}
   {\sigma_i^2} + {\cal O} \sum_i (c_i - m_i) 
\eeq
\beq
\frac{1}{2}\,\frac{\partial^2 \chi^2}{\partial g^{j'} \partial b^{k'}} 
 = \sum_i \frac{B_{k'i} G_{j'i} \left[\Delta T_{vi} + \Delta T_{ai}\right]}
   {\sigma_i^2}
\eeq
\beq
\frac{1}{2}\,\frac{\partial^2 \chi^2}{\partial a_{l'm'} \partial a_{l''m''}}
  = \sum_i \frac{g_i \left[Y_{l'm'}(\hat{n}_{Ai}) - Y_{l'm'}(\hat{n}_{Bi})\right]
              \, g_i \left[Y_{l''m''}(\hat{n}_{Ai}) - Y_{l''m''}(\hat{n}_{Bi})\right]}
    {\sigma_i^2}
\eeq
\beq
\frac{1}{2}\,\frac{\partial^2 \chi^2}{\partial a_{l'm'} \partial b^{k'}}
 = \sum_i \frac{g_i B_{k'i} \left[Y_{l'm'}(\hat{n}_{Ai}) - Y_{l'm'}(\hat{n}_{Bi})\right]}
   {\sigma_i^2}
\eeq
\beq
\frac{1}{2}\,\frac{\partial^2 \chi^2}{\partial b^{k'} \partial b^{k''}}
  = \sum_i \frac{B_{k''i} B_{k'i}}{\sigma_i^2}
\eeq

From this we can form the inverse covariance matrix
\beq
C^{-1} = \left(
\begin{array}{ccc}
 \frac{1}{2}\frac{\partial^2 \chi^2}{\partial g^{j'} \partial g^{j''}} & 
 \frac{1}{2}\frac{\partial^2 \chi^2}{\partial g^{j'} \partial a_{l''m''}} & 
 \frac{1}{2}\frac{\partial^2 \chi^2}{\partial g^{j'} \partial b^{k''}} 
 \vspace{3mm} \\
 \frac{1}{2}\frac{\partial^2 \chi^2}{\partial a_{l'm'} \partial g^{j''}} & 
 \frac{1}{2}\frac{\partial^2 \chi^2}{\partial a_{l'm'} \partial a_{l''m''}} & 
 \frac{1}{2}\frac{\partial^2 \chi^2}{\partial a_{l'm'} \partial b^{k''}} 
 \vspace{3mm} \\
 \frac{1}{2}\frac{\partial^2 \chi^2}{\partial b^{k'} \partial g^{j''}} & 
 \frac{1}{2}\frac{\partial^2 \chi^2}{\partial b^{k'} \partial a_{l''m''}} & 
 \frac{1}{2}\frac{\partial^2 \chi^2}{\partial b^{k'} \partial b^{k''}}
\end{array}
\right),
\eeq
where the gain and baseline blocks are $(2j_{\rm max}+1) \times (2j_{\rm
max}+1)$, and the sky map block is $(l_{\rm max}+1)^2 \times (l_{\rm max}+1)^2$.

If we decompose $C^{-1}$ using SVD the parameter covariance matrix can be
inverted to have the form
\beq
C = \sum_i \frac{1}{w_i} V_{(i)} \otimes V_{(i)}
\eeq
where the $w_i$ are the singular values, and the $V_{(i)}$ are the columns of
the orthogonal matrix $V$.  In this form, the uncertainty in the linear
combination of parameters defined by $V_{(i)}$ is $1/w_i$.

\section{CALIBRATION MODEL FITTING WITH GAIN ERROR TEMPLATES}
\label{app:cal_model_fit}

\subsection{Gain Error From Calibration Dipole Error}

Consider a simple model where the input sky consists of only a pure fixed (CMB)
dipole, described by the vector ${\bf d}_c$, and a dipole modulated by the
motion of \map\ with respect to the Sun, described by the time-dependent vector
${\bf d}_v(t)$.  The raw data produced by an experiment observing this signal is
\beq
c(t_i) = g(t_i)[\Delta t_c(t_i) + \Delta t_v(t_i)]
\eeq
where $c(t_i)$ is the TOD signal in counts, $g(t_i)$ is the true gain of the
instrument and $\Delta t_m(t_i)$ is the differential signal produced by each
dipole component ($m = c,v$) at time $t_i$ given the instrument pointing at that
time.  Note that we have suppressed the explicit baseline and noise terms here
for simplicity.

Now suppose we calibrate the instrument using an erroneous CMB dipole,  ${\bf
d}'_c = r{\bf d}_c = (1+\Delta r){\bf d}_c$, where $r$ is a number of order one
(and $\Delta r \ll 1$ so we can ignore terms of order $\Delta r^2$).  The fit
gain, $g_f(t)$, will then roughly have the form
\beq
g_f(t) = \frac{c(t)}{|{\bf d}'_c + {\bf d}_v(t)|}
    = g(t)\frac{|{\bf d}_c + {\bf d}_v(t)|}{|r{\bf d}_c + {\bf d}_v(t)|},
\eeq
where the vertical bars indicate vector magnitude.  Now define ${\bf d} \equiv
{\bf d}_c + {\bf d}_v$ and expand to 1st order in $\Delta r$ to get
\beq
g_f(t) = g(t)\left[1 - \Delta r \frac{{\bf d}(t)\cdot{\bf d}_c}{{\bf d}(t)\cdot{\bf d}(t)}\right].
\eeq
Note that the term $({\bf d}\cdot{\bf d}_c)/({\bf d}\cdot{\bf d})$ is dominated
by a constant component of order $d_c^2/(d_c^2+d_v^2) \sim 0.99$, followed by an
annually modulated term that is suppressed by a factor of order $d_v/d_c$.  Thus
an erroneous calibration dipole induces a specific error in the fit gain that
can be identified and corrected for, assuming the time dependence of the true
gain is orthogonal to this form.   

\subsection{Gain Model Fitting}

In theory, the way to do this is as follows.  We have a set of data in the form
of the fit gains, $g_{f,i}$ for each calibration sequence $i$, and we have a
gain model, $G(t;p_n)$, which is a function of time and a set of model
parameters $p_n$.  Ideally we would like to fit the model to the true gain,
$g(t)$, but since we don't know the true gain, the next best thing is to modify
the gain model to have the same modulation form as the dipole gains have and to
fit for this modulation simultaneously with the other gain model parameters. 
Thus $\chi^2$ takes the form
\beq
\chi^2 = \sum_i \frac{\left[g_i - G_i(p_n)\right]^2}{\sigma_i^2}
       = \sum_i \frac{\left[g_{f,i} - G_i(p_n)(1 - \Delta r f_{d,i})\right]^2}{\sigma_i^2},
\eeq
where $f_{d,i} \equiv ({\bf d}\cdot{\bf d}_c)/({\bf d}\cdot{\bf d})$ evaluated
at time $t_i$, or is a function generated from simulations.

Since the system is nonlinear, it must be minimized using a suitable nonlinear
least squares routine.  However, we can analyze the parameter covariance matrix
directly by explicitly evaluating the 2nd derivative of $\chi^2$ with respect to
the model parameters
\beq
C^{-1} = \frac{1}{2}\,\frac{\partial^2 \chi^2}{\partial p_j \partial p_k}.
\eeq
First compile the necessary 1st derivatives
\beq
\frac{1}{2}\,\frac{\partial \chi^2}{\partial \Delta r}
 = \sum_i \frac{\left[g_{f,i} - G_i(p_n)(1 - \Delta r f_{d,i})\right]\,
   (G_i \, f_{d,i})}
   {\sigma_i^2}
\eeq
\beq
\frac{1}{2}\,\frac{\partial \chi^2}{\partial p_m}
 = \sum_i \frac{\left[g_{f,i} - G_i(p_n)(1 - \Delta r f_{d,i})\right]\,
   (-\partial G_i / \partial p_m)(1 - \Delta r f_{d,i})}
   {\sigma_i^2}
\eeq
(We evaluate the individual $\partial G / \partial p_m$ terms below.)  Next the
various 2nd derivatives are
\beq
\frac{1}{2}\,\frac{\partial^2 \chi^2}{\partial \Delta r \partial \Delta r}
 = \sum_i \frac{(G_i \, f_{d,i})(G_i \, f_{d,i})}
   {\sigma_i^2},
\eeq
\beq
\frac{1}{2}\,\frac{\partial^2 \chi^2}{\partial \Delta r \partial p_m}
 = \sum_i \frac{(G_i \, f_{d,i})(-\partial G_i / \partial p_m)(1 - \Delta r f_{d,i})}
   {\sigma_i^2}
 + {\cal O} \sum_i (g_i - G_i),
\eeq
\beq
\frac{1}{2}\,\frac{\partial^2 \chi^2}{\partial p_m \partial p_n}
 = \sum_i \frac{(\partial G_i / \partial p_m)(1 - \Delta r f_{d,i})
                (\partial G_i / \partial p_n)(1 - \Delta r f_{d,i})}
   {\sigma_i^2}
 + {\cal O} \sum_i (g_i - G_i).
\eeq
In the last two expressions, we neglect the term proportional to $\partial^2G /
\partial p_m \partial p_n$ because the prefactor of $(g_i-G_i)$ is statistically
zero for the least squares solution.

Finally, we evaluate the $\partial G / \partial p_m$ terms.  The gain model has
the form \citep{jarosik/etal:2007}
\beq
G_i = \alpha\frac{\bar{V}(t_i)-V_0-\beta(T_{\rm RXB}(t_i)-T^0_{\rm RXB})}{T_{\rm FPA}(t_i)-T^0_{\rm FPA}},
\eeq
where $T^0_{\rm RXB} \equiv 290$ K, and $\alpha$, $V_0$, and $T^0_{\rm FPA}$ are
parameters to be fit.  The necessary 1st derivatives are
\beq
\partial G_i / \partial \alpha 
 = \frac{\bar{V}(t_i)-V_0-\beta(T_{\rm RXB}(t_i)-T^0_{\rm RXB})}{T_{\rm FPA}(t_i)-T^0_{\rm FPA}},
\eeq
\beq
\partial G_i / \partial V_0
 = \frac{-\alpha}{T_{\rm FPA}(t_i)-T^0_{\rm FPA}},
\eeq
\beq
\partial G_i / \partial \beta 
 = \frac{-\alpha(T_{\rm RXB}(t_i)-T^0_{\rm RXB})}{T_{\rm FPA}(t_i)-T^0_{\rm FPA}},
\eeq
\beq
\partial G_i / \partial T^0_{\rm FPA}
 = \alpha\frac{\bar{V}(t_i)-V_0-\beta(T_{\rm RXB}(t_i)-T^0_{\rm RXB})}
   {(T_{\rm FPA}(t_i)-T^0_{\rm FPA})^2}.
\eeq


\clearpage

\clearpage

\clearpage

\clearpage

\clearpage

\clearpage

\clearpage

\clearpage

\clearpage

\clearpage

\clearpage

\clearpage

\clearpage


\begin{deluxetable}{cccccccccc}
\tablecaption{Differencing Assembly (DA) Properties \label{tab:radiometers}}
\tablewidth{0pt}
\tablehead{
\colhead{DA} &
\colhead{$\lambda$\tablenotemark{a}} &
\colhead{$\nu$\tablenotemark{a}} &
\colhead{$g(\nu)$\tablenotemark{b}} &
\colhead{$\theta_{\rm FWHM}$\tablenotemark{c}} &
\colhead{$\sigma_0$(I)\tablenotemark{d}} &
\colhead{$\sigma_0$(Q,U)\tablenotemark{d}} &
\colhead{$\nu_{\rm s}$\tablenotemark{e}} &
\colhead{$\nu_{\rm ff}$\tablenotemark{e}} &
\colhead{$\nu_{\rm d}$\tablenotemark{e}} \\
\colhead{} &
\colhead{(mm)} &
\colhead{(GHz)} &
\colhead{} &
\colhead{($^\circ$)} &
\colhead{(mK)} &
\colhead{(mK)} &
\colhead{(GHz)} &
\colhead{(GHz)} &
\colhead{(GHz)}}
\startdata
 K1 & 13.17 & 22.77 & 1.0135 & 0.807 & 1.436 & 1.453 & 22.47 & 22.52 & 22.78 \\
Ka1 & 9.079 & 33.02 & 1.0285 & 0.624 & 1.470 & 1.488 & 32.71 & 32.76 & 33.02 \\
 Q1 & 7.342 & 40.83 & 1.0440 & 0.480 & 2.254 & 2.278 & 40.47 & 40.53 & 40.85 \\
 Q2 & 7.382 & 40.61 & 1.0435 & 0.475 & 2.141 & 2.163 & 40.27 & 40.32 & 40.62 \\
 V1 & 4.974 & 60.27 & 1.0980 & 0.324 & 3.314 & 3.341 & 59.65 & 59.74 & 60.29 \\
 V2 & 4.895 & 61.24 & 1.1010 & 0.328 & 2.953 & 2.975 & 60.60 & 60.70 & 61.27 \\
 W1 & 3.207 & 93.49 & 1.2480 & 0.213 & 5.899 & 5.929 & 92.68 & 92.82 & 93.59 \\
 W2 & 3.191 & 93.96 & 1.2505 & 0.196 & 6.565 & 6.602 & 93.34 & 93.44 & 94.03 \\
 W3 & 3.226 & 92.92 & 1.2445 & 0.196 & 6.926 & 6.964 & 92.34 & 92.44 & 92.98 \\
 W4 & 3.197 & 93.76 & 1.2495 & 0.210 & 6.761 & 6.800 & 93.04 & 93.17 & 93.84 \\
\enddata
\tablenotetext{a}{Effective wavelength and frequency for a thermodynamic
spectrum.}
\tablenotetext{b}{Conversion from antenna temperature to thermodynamic
temperature, $\Delta T = g(\nu) \Delta T_A$.}
\tablenotetext{c}{Full-width-at-half-maximum from radial profile of A- and
B-side average beams.  Note: beams are not Gaussian.}
\tablenotetext{d}{Noise per observation for resolution 9 and 10 $I$, $Q$, \& $U$
maps, to $\sim$0.1\% uncertainty. $\sigma(p)=\sigma_0 N_{\rm obs}^{-1/2}(p)$.}
\tablenotetext{e}{Effective frequency for synchrotron (s), free-free (ff), and
dust (d) emission, assuming spectral indices of $\beta = -2.9, -2.1, +2.0$,
respectively, in antenna temperature units.}
\end{deluxetable}

\begin{deluxetable}{lccccc}
\tablecaption{Lost and Rejected Data\label{tab:baddata}}
\tablewidth{0pt}
\tablehead{
\colhead{Category} & \colhead{K-band} & \colhead{Ka-band} & \colhead{Q-band} &
\colhead{V-band} & \colhead{W-band}}
\startdata
Lost or incomplete telemetry(\%)          & 0.12 & 0.12 & 0.12 & 0.12 & 0.12 \\
Spacecraft anomalies(\%)                  & 0.44 & 0.46 & 0.52 & 0.44 & 0.48 \\
Planned stationkeeping maneuvers(\%)      & 0.39 & 0.39 & 0.39 & 0.39 & 0.39 \\
Planet in beam (\%)                       & 0.11 & 0.11 & 0.11 & 0.11 & 0.11 \\
                                          &------&------&------&------&------\\
Total lost or rejected (\%)               & 1.06 & 1.08 & 1.14 & 1.06 & 1.10 \\
\enddata
\end{deluxetable}

\begin{deluxetable}{lrrrr}
\tablecaption{Change in low-$l$ Power from 3 year Data \label{tab:i_diff}}
\tablewidth{0pt}
\tablehead{
\colhead{Band} &
\colhead{$l=0$\tablenotemark{a}} &
\colhead{$l=1$\tablenotemark{a}} &
\colhead{$l=2$\tablenotemark{b}} &
\colhead{$l=3$\tablenotemark{b}} \\
\colhead{} &
\colhead{($\mu$K)} &
\colhead{($\mu$K)} &
\colhead{($\mu$K$^2$)} &
\colhead{($\mu$K$^2$)} }
\startdata
 K &  9.3 &  5.1 &  4.1 &  0.7 \\
Ka & 18.9 &  2.1 &  2.8 &  0.2 \\
 Q & 18.3 &  0.4 &  2.5 &  0.5 \\
 V & 14.4 &  7.3 &  1.2 &  0.0 \\
 W & 16.4 &  3.5 &  1.0 &  0.0 \\
\enddata
\tablenotetext{a}{$l=0,1$ - Amplitude in the difference map, outside the
processing cut, in $\mu$K.}
\tablenotetext{b}{$l=2,3$ - Power in the difference map, outside the processing
cut, $l(l+1)\,C_l/2\pi$, in $\mu$K$^2$.}
\end{deluxetable}

\begin{deluxetable}{lrrrrrr}
\tablecaption{\map\ 5 year CMB Dipole Anisotropy\tablenotemark{a}\label{tab:dipole}}
\tablewidth{0pt}
\tablehead{
\colhead{Cleaning} &
\colhead{$d_x$\tablenotemark{b}} &
\colhead{$d_y$} &
\colhead{$d_z$} &
\colhead{$d$\tablenotemark{c}} &
\colhead{$l$} &
\colhead{$b$} \\
\colhead{method} &
\colhead{(mK)} &
\colhead{(mK)} &
\colhead{(mK)} &
\colhead{(mK)} &
\colhead{($^{\circ}$)} &
\colhead{($^{\circ}$)} }
\startdata
Templates  & $-0.229 \pm 0.003$ & $-2.225 \pm 0.003$ & $2.506 \pm 0.003$ & $3.359 \pm 0.008$ & $264.11 \pm 0.08$ & $48.25 \pm 0.03$ \\
ILC        & $-0.238 \pm 0.003$ & $-2.218 \pm 0.002$ & $2.501 \pm 0.001$ & $3.352 \pm 0.007$ & $263.87 \pm 0.07$ & $48.26 \pm 0.02$ \\
Combined   & $-0.233 \pm 0.005$ & $-2.222 \pm 0.004$ & $2.504 \pm 0.003$ & $3.355 \pm 0.008$ & $263.99 \pm 0.14$ & $48.26 \pm 0.03$ \\
\enddata
\tablenotetext{a}{The CMB dipole components for two different galactic cleaning
methods are given in the first two rows.  The Gibbs samples from each set are
combined in the last row to produce an estimate with conservative uncertainties
that encompasses both cases.}
\tablenotetext{b}{The cartesian dipole components are given in Galactic
coordinates.  The quoted uncertainties reflect the effects of noise and sky cut,
for illustration.  An absolute calibration uncertainty of 0.2\% should be added
in quadrature.}
\tablenotetext{c}{The spherical components of the dipole are given in Galactic
coordinates.  In this case the quoted uncertainty in the magnitude, $d$,
includes the absolute calibration uncertainty.}
\end{deluxetable}

\begin{deluxetable}{lcccccccccc}
\tablecaption{Polarization $\chi^2$ Consistency Tests\tablenotemark{a}\label{tab:chi2_pol_l}}
\tablewidth{0pt}
\tablehead{\colhead{Multipole} & 
\colhead{KaKa} & \colhead{KaQ} & \colhead{KaV} & \colhead{KaW} & \colhead{QQ} &
\colhead{QV} & \colhead{QW} & \colhead{VV} & \colhead{VW} & \colhead{WW} \\
\colhead{} & 
\colhead{(10)\tablenotemark{b}} & \colhead{(50)} & \colhead{(50)} & \colhead{(100)} & \colhead{(45)} &
\colhead{(100)} & \colhead{(200)} & \colhead{(45)} & \colhead{(200)} & \colhead{(190)}}
\startdata
\multicolumn{11}{l}{EE} \\
 2 &  0.727 & 1.059 & 1.019 & 1.301 & 1.586 & 0.690 & 1.179 & 0.894 & 1.078 & 1.152 \\
 3 &  1.373 & 0.994 & 1.683 & 1.355 & 1.092 & 1.614 & 1.325 & 1.005 & 1.386 & 1.519 \\
 4 &  1.561 & 1.816 & 1.341 & 2.033 & 0.993 & 1.126 & 1.581 & 1.195 & 1.596 & 1.724 \\
 5 &  0.914 & 1.313 & 1.062 & 1.275 & 1.631 & 1.052 & 1.155 & 0.589 & 0.881 & 1.252 \\
 6 &  1.003 & 0.847 & 0.688 & 1.124 & 0.740 & 0.856 & 1.049 & 1.384 & 1.168 & 1.142 \\
 7 &  0.600 & 0.671 & 0.689 & 0.936 & 0.936 & 0.780 & 0.864 & 0.900 & 1.064 & 1.015 \\
 8 &  1.578 & 1.262 & 1.337 & 1.212 & 1.080 & 0.763 & 0.608 & 1.025 & 0.871 & 0.749 \\
 9 &  0.760 & 0.710 & 0.891 & 0.820 & 0.582 & 0.726 & 0.651 & 0.791 & 0.821 & 0.795 \\
10 &  0.494 & 0.821 & 0.996 & 0.914 & 0.656 & 0.763 & 0.806 & 0.676 & 0.891 & 0.943 \\[2.0mm]
\multicolumn{11}{l}{EB} \\
 2 &  0.900 & 1.297 & 1.179 & 2.074 & 1.006 & 0.915 & 2.126 & 1.242 & 2.085 & 2.309 \\
 3 &  0.719 & 1.599 & 0.651 & 2.182 & 1.295 & 0.986 & 2.739 & 1.095 & 3.276 & 3.157 \\
 4 &  0.746 & 1.702 & 1.378 & 1.777 & 1.926 & 1.110 & 1.435 & 1.028 & 1.279 & 1.861 \\
 5 &  1.161 & 0.948 & 0.945 & 1.003 & 1.149 & 1.232 & 1.468 & 0.699 & 1.122 & 1.516 \\
 6 &  0.475 & 1.183 & 0.651 & 0.687 & 0.829 & 1.023 & 0.814 & 1.201 & 1.136 & 0.960 \\
 7 &  1.014 & 1.007 & 0.829 & 0.700 & 0.817 & 0.759 & 1.112 & 0.616 & 0.802 & 1.233 \\
 8 &  0.849 & 0.897 & 1.279 & 0.861 & 0.681 & 0.689 & 0.955 & 1.021 & 0.954 & 0.996 \\
 9 &  0.743 & 0.734 & 1.007 & 1.112 & 0.820 & 0.798 & 0.686 & 0.882 & 0.808 & 0.824 \\
10 &  0.413 & 1.003 & 1.316 & 0.859 & 0.722 & 0.900 & 0.693 & 1.124 & 0.836 & 0.852 \\[2.0mm]
\multicolumn{11}{l}{BB} \\
 2 &  2.038 & 1.570 & 1.244 & 2.497 & 1.340 & 1.219 & 2.529 & 0.694 & 1.631 & 9.195 \\
 3 &  0.756 & 0.868 & 0.808 & 1.817 & 3.027 & 1.717 & 3.496 & 0.601 & 2.545 & 5.997 \\
 4 &  1.058 & 1.455 & 1.522 & 2.144 & 1.007 & 0.905 & 1.786 & 0.752 & 1.403 & 1.984 \\
 5 &  1.221 & 1.659 & 1.742 & 2.036 & 0.889 & 1.057 & 1.271 & 1.078 & 1.660 & 1.255 \\
 6 &  0.379 & 0.805 & 0.483 & 0.812 & 1.009 & 0.861 & 1.238 & 0.800 & 0.767 & 0.955 \\
 7 &  1.925 & 1.594 & 0.967 & 1.332 & 1.074 & 0.817 & 0.928 & 0.772 & 0.994 & 1.024 \\
 8 &  0.804 & 1.005 & 0.999 & 0.912 & 1.069 & 0.782 & 0.831 & 0.997 & 0.879 & 0.943 \\
 9 &  0.320 & 0.489 & 0.502 & 0.450 & 0.884 & 0.491 & 0.729 & 0.748 & 0.664 & 0.959 \\
10 &  1.181 & 1.162 & 1.028 & 0.980 & 1.218 & 1.165 & 0.951 & 1.079 & 0.621 & 0.791 \\
\enddata
\tablenotetext{a}{Table gives $\chi^2$ per degree of freedom of the independent
spectrum estimates per multipole per band or band-pair, estimated from the
template-cleaned maps.  See text for details.}
\tablenotetext{b}{Second header row indicates the number of degrees of freedom
in the reduced $\chi^2$ for that spectrum.  See text for details.}
\end{deluxetable}

\begin{deluxetable}{lrr}
\tablecaption{Loss Imbalance Coefficients\tablenotemark{a}}
\tablewidth{0pt}
\tablehead{
\colhead{DA} & \colhead{$x_{\rm im,1}$} & \colhead{$x_{\rm im,2}$} \\
\colhead{} & \colhead{(\%)} & \colhead{(\%)} }
\startdata
K1  &  0.012 & 0.589 \\
Ka1 &  0.359 & 0.148 \\
Q1  & -0.031 & 0.412 \\
Q2  &  0.691 & 1.048 \\
V1  &  0.041 & 0.226 \\
V2  &  0.404 & 0.409 \\
W1  &  0.939 & 0.128 \\
W2  &  0.601 & 1.140 \\
W3  & -0.009 & 0.497 \\
W4  &  2.615 & 1.946 \\
\enddata
\tablenotetext{a}{{Loss} imbalance is defined as $x_{\rm im} = (\epsilon_A -
\epsilon_B)/(\epsilon_A + \epsilon_B)$.  See \S\ref{sec:pol_emiss} and
\citet{jarosik/etal:2007} for details.}
\label{tab:x_im}
\end{deluxetable}

\begin{deluxetable}{lccc}
\tabletypesize{\footnotesize}
\tablecaption{Cosmological Parameter Summary \label{tab:best_param}}
\tablewidth{0pt}
\tablehead{
\colhead{Description} & \colhead{Symbol} & \colhead{\map-only} & \colhead{\map+BAO+SN}}
\startdata
\multicolumn{4}{c}{Parameters for Standard $\Lambda$CDM Model \tablenotemark{a}} 
    \\[3.0mm]
Age of universe
    &\ensuremath{t_0}
    &\ensuremath{13.69\pm 0.13\ \mbox{Gyr}}
    &\ensuremath{13.72\pm 0.12\ \mbox{Gyr}}
    \\[1.5mm]
Hubble constant
    &\ensuremath{H_0}
    &\ensuremath{71.9^{+ 2.6}_{- 2.7}\ \mbox{km/s/Mpc}}
    &\ensuremath{70.5\pm 1.3\ \mbox{km/s/Mpc}}
    \\[1.5mm]
Baryon density
    &\ensuremath{\Omega_b}
    &\ensuremath{0.0441\pm 0.0030}
    &\ensuremath{0.0456\pm 0.0015}   
    \\[1.5mm]
Physical baryon density
    &\ensuremath{\Omega_bh^2}
    &\ensuremath{0.02273\pm 0.00062}
    &\ensuremath{0.02267^{+ 0.00058}_{- 0.00059}}
    \\[1.5mm]
Dark matter density
    &\ensuremath{\Omega_c}
    &\ensuremath{0.214\pm 0.027}
    &\ensuremath{0.228\pm 0.013}
    \\[1.5mm]
Physical dark matter density
    &\ensuremath{\Omega_ch^2}
    &\ensuremath{0.1099\pm 0.0062}
    &\ensuremath{0.1131\pm 0.0034}
    \\[1.5mm]
Dark energy density
    &\ensuremath{\Omega_\Lambda}
    &\ensuremath{0.742\pm 0.030}
    &\ensuremath{0.726\pm 0.015}
    \\[1.5mm]
Curvature fluctuation amplitude, $k_0=0.002$ Mpc$^{-1}$ \tablenotemark{b}
    &\ensuremath{\Delta_{\cal R}^2}
    &\ensuremath{(2.41\pm 0.11)\times 10^{-9}}
    &\ensuremath{(2.445\pm 0.096)\times 10^{-9}}
    \\[1.5mm]
Fluctuation amplitude at $8h^{-1}$ Mpc
    &\ensuremath{\sigma_8}
    &\ensuremath{0.796\pm 0.036}
    &\ensuremath{0.812\pm 0.026}
    \\[1.5mm]
$l(l+1)C^{TT}_{220}/2\pi$
    &\ensuremath{C_{220}}
    &\ensuremath{5756\pm 42} $\mu$K$^2$
    &\ensuremath{5751^{+ 42}_{- 43}} $\mu$K$^2$
    \\[1.5mm]
Scalar spectral index
    &\ensuremath{n_s}
    &\ensuremath{0.963^{+ 0.014}_{- 0.015}}
    &\ensuremath{0.960\pm 0.013}
    \\[1.5mm]
Redshift of matter-radiation equality
    &\ensuremath{z_{\rm eq}}
    &\ensuremath{3176^{+ 151}_{- 150}}  
    &\ensuremath{3253^{+ 89}_{- 87}}    
    \\[1.5mm]
Angular diameter distance to matter-radiation eq.\tablenotemark{c}
    &\ensuremath{d_A(z_{\rm eq})}
    &\ensuremath{14279^{+ 186}_{- 189}\ \mbox{Mpc}}
    &\ensuremath{14200^{+ 137}_{- 140}\ \mbox{Mpc}}
    \\[1.5mm]
Redshift of decoupling
    &\ensuremath{z_{*}}
    &\ensuremath{1090.51\pm 0.95}
    &\ensuremath{1090.88\pm 0.72}
    \\[1.5mm]
Age at decoupling
    &\ensuremath{t_{*}}
    &\ensuremath{380081^{+ 5843}_{- 5841}\ \mbox{yr}}
    &\ensuremath{376971^{+ 3162}_{- 3167}\ \mbox{yr}}
    \\[1.5mm]
Angular diameter distance to decoupling \tablenotemark{c,d}
    &\ensuremath{d_A(z_{*})}
    &\ensuremath{14115^{+ 188}_{- 191}\ \mbox{Mpc}}
    &\ensuremath{14034^{+ 138}_{- 142}\ \mbox{Mpc}}
    \\[1.5mm]
Sound horizon at decoupling \tablenotemark{d}
    &\ensuremath{r_s(z_*)}
    &\ensuremath{146.8\pm 1.8\ \mbox{Mpc}}
    &\ensuremath{145.9^{+ 1.1}_{- 1.2}\ \mbox{Mpc}}
    \\[1.5mm]
Acoustic scale at decoupling \tablenotemark{d}
    &$l_A(z_*)$
    &\ensuremath{302.08^{+ 0.83}_{- 0.84}}
    &\ensuremath{302.13\pm 0.84}
    \\[1.5mm]
Reionization optical depth
    &\ensuremath{\tau}
    &\ensuremath{0.087\pm 0.017}  
    &\ensuremath{0.084\pm 0.016}    
    \\[1.5mm]
Redshift of reionization
    &\ensuremath{z_{\rm reion}}
    &\ensuremath{11.0\pm 1.4}  
    &\ensuremath{10.9\pm 1.4}    
    \\[1.5mm]
Age at reionization
    &\ensuremath{t_{\rm reion}}
    &$427^{+88}_{-65}$ Myr
    &$432^{+90}_{-67}$ Myr
    \\[1.5mm]
\multicolumn{4}{c}{Parameters for Extended Models \tablenotemark{e}} 
    \\[3.0mm]
Total density \tablenotemark{f}
    &\ensuremath{\Omega_{\rm tot}}
    &\ensuremath{1.099^{+ 0.100}_{- 0.085}}
    &\ensuremath{1.0050^{+ 0.0060}_{- 0.0061}}
    \\[1.5mm]
Equation of state \tablenotemark{g}
    &\ensuremath{w}
    &\ensuremath{-1.06^{+ 0.41}_{- 0.42}}
    &\ensuremath{-0.992^{+ 0.061}_{- 0.062}}
    \\[1.5mm]
Tensor to scalar ratio, $k_0=0.002$ Mpc$^{-1}$ \tablenotemark{b,h}
    &\ensuremath{r}
    &\ensuremath{< 0.43\ \mbox{(95\% CL)}}
    &\ensuremath{< 0.22\ \mbox{(95\% CL)}}
    \\[1.5mm]
Running of spectral index, $k_0=0.002$ Mpc$^{-1}$ \tablenotemark{b,i}
    &\ensuremath{dn_s/d\ln{k}}
    &\ensuremath{-0.037\pm 0.028}
    &\ensuremath{-0.028\pm 0.020}
    \\[1.5mm]
Neutrino density \tablenotemark{j}
    &\ensuremath{\Omega_\nu h^2}
    &\ensuremath{< 0.014\ \mbox{(95\% CL)}}
    &\ensuremath{< 0.0071\ \mbox{(95\% CL)}}
    \\[1.5mm]
Neutrino mass \tablenotemark{j}
    &\ensuremath{\sum m_\nu}
    &\ensuremath{< 1.3\ \mbox{eV}\ \mbox{(95\% CL)}}
    &\ensuremath{< 0.67\ \mbox{eV}\ \mbox{(95\% CL)}}
    \\[1.5mm]
Number of light neutrino families \tablenotemark{k}
    &\ensuremath{N_{\rm eff}}
    &\ensuremath{> 2.3\ \mbox{(95\% CL)}}
    &\ensuremath{4.4\pm 1.5}
    \\[1.5mm]
\enddata
\tablenotetext{a}{The parameters reported in the first section assume the 6
parameter $\Lambda$CDM model, first using \map\ data only
\citep{dunkley/etal:prep}, then using \map+BAO+SN data
\citep{komatsu/etal:prep}.}
\tablenotetext{b}{$k=0.002$ Mpc$^{-1}$  $\longleftrightarrow$  $l_{\rm eff} \approx 30$.}
\tablenotetext{c}{Comoving angular diameter distance.}
\tablenotetext{d}{$l_A(z_*) \equiv \pi \, \ensuremath{d_A(z_{*})} \, \ensuremath{r_s(z_*)}^{-1}$.}
\tablenotetext{e}{The parameters reported in the second section place limits on
deviations from the $\Lambda$CDM model, first using \map\ data only
\citep{dunkley/etal:prep}, then using \map+BAO+SN data
\citep{komatsu/etal:prep}.  A complete listing of all parameter values and
uncertainties for each of the extended models studied is available on LAMBDA.}
\tablenotetext{f}{Allows non-zero curvature, $\Omega_k \ne 0$.}
\tablenotetext{g}{Allows $w \ne -1$, but assumes $w$ is constant.}
\tablenotetext{h}{Allows tensors modes but no running in scalar spectral index.}
\tablenotetext{i}{Allows running in scalar spectral index but no tensor modes.}
\tablenotetext{j}{Allows a massive neutrino component, $\Omega_{\nu} \ne 0$.}
\tablenotetext{k}{Allows $N_{\rm eff}$ number of relativistic species.  The last
column adds the HST prior to the other data sets.}
\end{deluxetable}

\end{document}